\documentclass[times,sort&compress,3p]{elsarticle}
\usepackage{amsmath, amssymb, enumerate, amsthm}
\usepackage[latin1]{inputenc}
\usepackage{times}
\usepackage{epsfig}
\usepackage{dsfont}
\usepackage{amsfonts}
\usepackage{float}
\usepackage{amsmath}
\usepackage{latexsym}
\usepackage{amssymb}
\usepackage{xr} 
\usepackage{natbib}
\usepackage{color}
\usepackage{multirow}
\usepackage{rotating}

\relax 
\allowdisplaybreaks[3]

\newcommand{\renu}{\mathbb{R}}

\newcommand{\1}{{\mathds{1}}}

\newcommand{\bqa}{\begin{eqnarray*}}
\newcommand{\eqa}{\end{eqnarray*}}

\newcommand{\bqan}{\begin{eqnarray}}
\newcommand{\eqan}{\end{eqnarray}}

\newcommand{\bit}{\begin{itemize}}
\newcommand{\eit}{\end{itemize}}

\newtheorem{definition}{{\sc Definition}\sc}[section]
\newcommand{\bdefi}{\begin{definition}}
	\newcommand{\edefi}{\end{definition}}

\newtheorem{theorem}[definition]{{\sc Theorem}\sc}

\newcommand{\ep}{\epsilon}

\newcommand{\ks}{\oplus}
\newcommand{\kp}{\otimes}

\DeclareMathOperator{\rank}{rank}
\DeclareMathOperator{\Cov}{Cov}
\DeclareMathOperator{\Var}{Var}
\DeclareMathOperator{\tr}{tr}
\DeclareMathOperator{\diag}{diag}

\newcommand{\hnu}{{\widehat \nu}}

\newcommand{\hsigma}{{\widehat \sigma}}

\newcommand{\vDh}{\widehat{\boldsymbol{D}}}

\newcommand{\vVh}{\widehat{\boldsymbol{V}}}

\newcommand{\vSigmah}{ \widehat{\vSigma}}

\newcommand{\olX}{\overline{X}}
\newcommand{\olY}{\overline{Y}}

\newcommand{\ve}{\boldsymbol{e}}

\newcommand{\vh}{\boldsymbol{h}}

\newcommand{\vn}{\boldsymbol{n}}

\newcommand{\vD}{\boldsymbol{D}}

\newcommand{\vH}{\boldsymbol{H}}
\newcommand{\vI}{\boldsymbol{I}}
\newcommand{\vJ}{\boldsymbol{J}}

\newcommand{\vM}{\boldsymbol{M}}

\newcommand{\vP}{\boldsymbol{P}}
\newcommand{\vQ}{\boldsymbol{Q}}

\newcommand{\vT}{\boldsymbol{T}}

\newcommand{\vV}{\boldsymbol{V}}

\newcommand{\vX}{\boldsymbol{X}}
\newcommand{\vY}{\boldsymbol{Y}}

\newcommand{\vep}{\boldsymbol{\epsilon}}

\newcommand{\vlam}{\boldsymbol{\lambda}}

\newcommand{\vmu}{\boldsymbol{\mu}}
\newcommand{\vnu}{\boldsymbol{\nu}}

\newcommand{\vSigma}{\boldsymbol{\Sigma}}

\newcommand{\vOmega}{\boldsymbol{\Omega}}

\newcommand{\veins}{{\bf 1}}
\newcommand{\vnull}{{\bf 0}}

\newcommand{\volX}{\boldsymbol{\overline{X}}}
\newcommand{\volY}{\boldsymbol{\overline{Y}}}

\newcommand{\volmu}{\boldsymbol{\overline{\mu}}}

\newcommand{\vwhe}{\boldsymbol{\widehat{e}}}

\newcommand{\vwhD}{\boldsymbol{\widehat{D}}}

\newcommand{\vwhV}{\boldsymbol{\widehat{V}}}

\newcommand{\vwhSigma}{\boldsymbol{\widehat{\Sigma}}}

\newcommand{\vwtV}{\boldsymbol{\widetilde{V}}}

\newcommand{\vwtX}{\boldsymbol{\widetilde{X}}}
\newcommand{\vwtY}{\boldsymbol{\widetilde{Y}}}

\newcommand{\vwtSigma}{\boldsymbol{\widetilde{\Sigma}}}

\newcommand{\E}{{\rm E}}

\numberwithin{equation}{section}

\begin{document}



\title{MATS: Inference for potentially Singular and Heteroscedastic MANOVA}

\author{Sarah Friedrich\corref{cor1}}
\ead{sarah.friedrich@uni-ulm.de}
\author{Markus Pauly}

\address{ Institute of Statistics, Ulm University, Germany}

\begin{abstract}
	In many experiments in the life sciences, several endpoints
	are recorded per subject. 
	The analysis of such multivariate data is usually based on MANOVA models assuming multivariate normality and covariance homogeneity.
	These assumptions, however, are often not met in practice. Furthermore, test statistics should be invariant under scale transformations of the data, since the endpoints may be measured on different scales.
	In the context of high-dimensional data, Srivastava and Kubokawa (2013) proposed such a test statistic for a specific one-way model, which, however, relies on the assumption of a common non-singular covariance matrix.
	We modify and extend this test statistic to factorial MANOVA designs, incorporating general heteroscedastic models. In particular, our only distributional assumption is the existence of the group-wise covariance matrices, which may even be singular. 
	We base inference on quantiles of resampling distributions, and derive confidence regions and ellipsoids based on these quantiles. 
	In a simulation study, we extensively analyze the behavior of these procedures. 
	Finally, the methods are applied to a {\color{black} data set containing information on the 2016 presidential elections in the USA with }unequal and singular {\color{black}empirical} covariance matrices.
\end{abstract}




\maketitle

\newpage


\section{Motivation and Introduction}\label{int}

In many experiments in{\color{black}, e.g., biology, ecology and psychology} several endpoints, potentially measured on different scales, are recorded per subject. 
{\color{black} As an example, we consider a data set on the 2016 presidential elections in the USA 
containing demographic data on counties from US census. For our exemplary 
analysis, we aim to investigate whether the states differ with respect to some demographic factors.}
In addition to unequal empirical covariance matrices between groups, analysis is further complicated by their singularity.\\
The analysis of such multivariate data is typically based on classical MANOVA models assuming multivariate normality and/or homogeneity of the covariance matrices, see, e.g., \cite{bartlett, Dempster1958, Dempster1960, Hotelling:1951, Lawley, Pillai, Wilks1946}. These assumptions, however, are often not met in practice (as in the motivating example) and it is well known that the methods perform poorly in case of heterogeneous data~\citep{Kon:2015, VallejoAto2012}. Furthermore, the test statistic should be invariant under scale transformations of the components, since the endpoints may be measured on different scales.
Thus, test statistics of multivariate ANOVA-type (ATS) as, e.g., proposed in \cite{brunner:2001} and studied in \cite{friedrich2017permuting}, are only applicable if all endpoints are measured on the same scale, i.e., for repeated measures designs. Assuming non-singular covariance matrices and certain moment assumptions the scale invariance is typically accomplished by utilizing test statistics of Wald-type (WTS). However, inference procedures based thereon require (extremely) large sample sizes for being accurate, see \cite{Kon:2015, smaga2016bootstrap, Vallejo}. In particular, even the novel approaches of \cite{Kon:2015} and \cite{smaga2016bootstrap} showed a more or less liberal behavior in case of skewed distributions. Moreover, their procedures cannot be used to analyze the motivating data example with possibly singular covariance matrices. Therefore, we follow a different approach by modifying the above mentioned ANOVA-type statistic (MATS). It is motivated from the modified Dempster statistic proposed in \cite{srivastava2013tests} for high-dimensional one-way MANOVA. This statistic is also invariant under the change of units of measurements. However, until now, it has only been developed for a homoscedastic one-way setting assuming non-singularity and a specific distributional structure that is motivated from multivariate normality.

It is the aim of the present paper to modify and extend the \cite{srivastava2013tests} approach to factorial MANOVA designs, incorporating general heteroscedastic models. In particular, we only postulate existence of the group-wise covariance matrices, which may even be singular. 
The small sample behavior of our test statistic is enhanced by applying bootstrap techniques as in \cite{Kon:2015}. Thereby, the novel MATS procedure enables us to relax the usual MANOVA assumptions in several ways: While incorporating general heteroscedastic designs and allowing for potentially singular covariance matrices we postulate their existence as solely distributional assumption, i.e., only finite second moments are required. Moreover, we gain a procedure that is more robust against deviations from symmetry and homoscedasticity than the usual WTS approaches.

So far, only few approaches have been investigated which do not assume normality or equal covariance matrices (or both). Examples in the nonparametric framework include the permutation based nonparametric combination method \citep{pesarin-salmaso-book,Pesarin:2012} and the rank-based tests presented in 
\cite{brunner2016rank} and \cite{brunner:99} for Split Plot Designs and in \cite{bathke2008compare} and \cite{liu2011nonparametric} for MANOVA designs. However, these methods are either not applicable for general MANOVA models or based on null hypotheses formulated in terms of distribution functions. In contrast we wish to derive inference procedures (tests and confidence regions) for contrasts of mean vectors. Here, beside all previously mentioned procedures, only methods for specific designs have been developed, see \cite{ChungRomano:2013} for two-sample problems, \cite{van2011robust, van2013fast} for robust but homoscedastic one-way MANOVA or \cite{harrar2012modified} for a particular two-way MANOVA. To our knowledge, mean-based MANOVA procedures allowing for possibly singular covariance matrices have not been developed so far.

The paper is organized as follows:
In Section \ref{mod} we describe the statistical model and hypotheses. Furthermore, we propose a new test statistic, which is applicable to singular covariance matrices and is invariant under scale transformations of the data. In Section \ref{sec:bs}, we present {\color{black}three} different resampling approaches, which are used for the derivation of statistical tests as well as confidence regions and simultaneous confidence intervals for contrasts in Section \ref{sec:stats}. The different approaches are compared in a large simulation study (Section \ref{sim}), where we analyze different factorial designs with a wide variety of
distributions and covariance settings. The motivating data example is analyzed in Section \ref{app} and we conclude with some final remarks and discussion in Section \ref{dis}. All proofs are deferred to the supplementary material, where we also provide further simulation results {\color{black} and the analysis of an additional data example}.

\section{Statistical Model, Hypotheses and Test Statistics} \label{mod}

Throughout the paper, we will use the following notation. We denote by $\vI_d$ the $d$-dimensional unit matrix and by $\vJ_d$ the $d \times d$ matrix of 1's, i.e., $\vJ_d = \boldsymbol{1}_d \boldsymbol{1}_d^\top$, where $\boldsymbol{1}_d=(1, \dots, 1)^\top$
is the $d$-dimensional column vector of 1's. Furthermore, let $\vP_d = \vI_d - d^{-1} \vJ_d$ denote the $d$-dimensional centering matrix. By $\ks$ and $\kp$ we denote the direct sum and the Kronecker product, respectively.

In order to cover different factorial designs of interest, we establish the general model
\bqa
\vX_{ik}&= \vmu_i + \vep_{ik}
\eqa
for treatment group $i=1, \dots, a$ and individual $k=1, \dots,n_i$, on which we measure $d$-variate observations. Here $\vmu_i = (\mu_{i1}, \dots, \mu_{id})^\top \in \mathbb{R}^d$ for $i=1, \dots, a$. A factorial structure can be incorporated by splitting up indices, see, e.g., \cite{Kon:2015}. For fixed $1\leq i \leq a$, the error terms $\vep_{ik}$ are assumed to be independent and identically distributed $d$-dimensional random vectors, for which the following conditions hold:
\begin{center}
\bit
\item[(1)] $\E(\vep_{i1}) = 0,~ i=1, \dots, a,$
\item[(2)] $0 < \sigma_{is}^2 = \Var(X_{iks}) < \infty ,~ i=1, \dots, a, ~ s=1, \dots, d,$
\item[(3)] $\Cov(\vep_{i1})=\vV_i \geq 0,~ i=1, \dots, a.$
\eit
\end{center}

Thus, we only assume the existence of second moments.
For convenience, we aggregate the individual vectors into $\vX=(\vX_{11}^\top, \dots, \vX_{an_a}^\top)^\top$ as well as $\vmu = (\vmu_{1}^\top, \dots, \vmu_a^\top)^\top$. Denote by $N = \sum_{i=1}^a n_i$ the total sample size. In order to derive asymptotic results in this framework, we will throughout assume the usual sample size condition:
\bqa
n_i/N \to \kappa_i > 0, i=1, \dots, a
\eqa
as $N \to \infty$.

An estimator for $\vmu$ is given by the vector of pooled group means $\volX_{i\cdot} = n_i^{-1}\sum_{k=1}^{n_i}\vX_{ik}, i=1, \dots, a$, which we denote by $\volX_{\bullet} = (\volX_{1\cdot}^\top, \dots, \volX_{a\cdot}^\top)^\top$. The covariance matrix of $\sqrt{N}~\volX_{\bullet}$ is given by 
\bqa
\vSigma_N = \Cov(\sqrt{N}~\volX_{\bullet}) = \diag \left(\frac{N}{n_i}\vV_i: 1 \leq i \leq a \right),
\eqa
where the group-specific covariance matrices $\vV_i$ are estimated by the empirical covariance matrices
\bqa
\vVh_i = \frac{1}{n_i-1}\sum_{k=1}^{n_i}(\vX_{ik} - \volX_{i \cdot})(\vX_{ik} - \volX_{i \cdot})^\top
\eqa
resulting in
\bqa
\vSigmah_N = \diag\left(\frac{N}{n_i}\vVh_i: 1 \leq i \leq a\right).
\eqa

In this semi-parametric framework, hypotheses are formulated in terms of the mean vector as $H_0: \vH \vmu = \vnull$, where $\vH$ is a suitable contrast matrix, i.e., $\vH \veins_{ad}=\vnull$.
Note that we can use the unique projection matrix $\vT = \vH^\top(\vH \vH^\top)^+\vH$, where $(\vH \vH^\top)^+$ denotes the Moore-Penrose inverse of $\vH \vH^\top$. It is $\vT = \vT^2, \vT = \vT^\top$ and $\vT \vmu = \vnull \Leftrightarrow \vH\vmu=\vnull$, see, e.g., \cite{brunnerPuri}.

A commonly used test statistic for multivariate data is the Wald-type statistic (WTS)
\bqan \label{WTS}
T_N =  N \volX_{\bullet}^\top \vT (\vT \vSigmah_N \vT)^+\vT \volX_{\bullet},
\eqan
which requires the additional assumption (3') $\vV_i > 0,~ i=1, \dots, a$.
It is easy to show that the WTS has under $H_0: \vT \vmu = \vnull$, asymptotically, as $N \to \infty$, a $\chi^2_f$-distribution with $f = \rank(\vT)$ degrees of freedom if (1) -- (3') holds. However, large sample sizes are necessary to maintain a pre-assigned level $\alpha$ using quantiles of the limiting $\chi^2$-distribution. \cite{Kon:2015} proposed different resampling procedures in order to improve the small sample behavior of the WTS for multivariate data. 
Therein, a parametric bootstrap approach turned out to be the best in case that the underlying distributions 
are not too skewed and/or too heteroscedastic. In the latter cases all considered procedures were more or less liberal. Moreover, assuming only (3) instead of (3') for the WTS would in general not lead to an asymptotic $\chi^2_f$-limit distribution. {\color{black} The reason for this are possible rank jumps between $\vT \vSigmah_N \vT$, $\vT \vSigma \vT$ and $\vT$. 
 To accept this, suppose that $\rank(\vT \vSigma \vT) = 2$, while $\rank(\vT) = 4$ (this corresponds to Scenario S5 in the simulation studies below). 
  If additionally $\lim_{N\to \infty} \rank(\vT \vSigmah_N \vT) = \rank(\vT \vSigma \vT) = 2$, we have that the WTS follows, asymptotically, a $\chi^2_{g}$-distribution under the null hypothesis, where $g=\rank(\vT \vSigma \vT)=2$. The Wald-type test, however, compares $T_N$ to the quantile of a $\chi^2_4$-distribution. Thus, for a chosen significance level of $\alpha=0.05$ this results in a true asymptotic $(N\to \infty)$ type-I error of 
$$
\Pr(T_N > \chi^2_{4; 0.95}) = 1 - \Pr(T_N \leq \chi^2_{4; 0.95}) \approx 0.0087, 
$$
i.e., a strictly conservative behavior of the test.  Here $\chi^2_{f; 1 - \alpha}$ denotes the $(1-\alpha)$-quantile of the $\chi^2_f$-distribution. Similarly, for $\alpha=0.1$ and $\alpha=0.01$ we obtain asymptotically inflated type-I error rates of 
$0.02$ and $0.0013$ (both again conservative), respectively. Moreover, the situation is even more complicated since $\lim_{N\to \infty} \rank(\vT \vSigmah_N \vT) = \rank(\vT \vSigma \vT)$ is neither verifiable in practice nor holds in general.

We tackle this problem in the current paper, where we not only} 
 relax the assumption (3') on the unknown covariance matrices but also gain a procedure that is more robust to deviations from symmetry and homoscedasticity. 
To this end, we consider a different test statistic, 
namely a multivariate version of the ANOVA-type statistic (ATS) proposed by \cite{brunner:2001} for repeated measures designs, 
which we obtain by erasing the Moore-Penrose term from \eqref{WTS}:
\bqa
\tilde{Q}_N = N \volX_{\bullet}^\top \vT \volX_{\bullet}.
\eqa
In the special two-sample case where we wish to test the null hypothesis $H_0: \vmu_1 - \vmu_2 = \vnull$, this is equivalent to the test statistic proposed by \cite{Dempster1960}.

The drawback of the ATS for  multivariate data is that it is not invariant under scale transformations of the components, 
e.g., under change of units ($cm \mapsto m$ or $g \mapsto kg$) in one or more components. {\color{black} We demonstrate this problem in a real data analysis given in the supplementary material, where we exemplify that a simple unit change can completely alter the test decision of the ATS. Thus, we} 
 consider a slightly modified version of the ATS, which we denote as MATS:
\bqan \label{modATS}
Q_N = N \volX_{\bullet}^\top \vT (\vT \vwhD_N \vT)^+\vT \volX_{\bullet}.
\eqan
Here, $\vwhD_N = \diag\left(N/n_i \cdot\hsigma_{is}^2 \right), i=1, \dots, a, ~ s=1, \dots, d$, where 
$\hsigma_{is}^2$ is the empirical variance of component $s$ in group $i$. 
A related test statistic has been proposed by 
\cite{srivastava2013tests} in the context of high-dimensional ($d\to \infty$) data for a special non-singular one-way MANOVA design. 
Here, we investigate in the classical multivariate case (with fixed $d$) 
how its finite sample performance can be enhanced considerably. We start by analyzing its asymptotic limit behavior.

\begin{theorem}\label{theo:MATS}
Under {\color{black} Conditions (1), (2) and (3) and under }$H_0: \vT \vmu = \vnull$, the test statistic $Q_N$ in \eqref{modATS} has asymptotically, as $N \to \infty$, the same distribution as a weighted sum of $\chi^2_1$ distributed random variables, where the weights $\lambda_{is}$ are the eigenvalues of $\vV = \vT (\vT \vD \vT)^+ \vT \vSigma$ for $\vD = \diag\left(\kappa_i^{-1} \sigma_{is}^2\right)$ and $\vSigma = \diag(\kappa_i^{-1} \vV_i)$, i.e.,
$$
Q_N = N \volX_{\bullet}^\top \vT (\vT \vwhD_N\vT)^+\vT \volX_{\bullet} \xrightarrow{\mathcal{D}} Z = \sum_{i=1}^a \sum_{s=1}^d \lambda_{is} Z_{is},
$$
with $Z_{is} \stackrel{\text{i.i.d.}}{\sim} \chi^2_1$ and "$\xrightarrow{\mathcal{D}}$" denoting convergence in distribution. 
\\
\end{theorem}

{\color{black}
Thus, we obtain an asymptotic level $\alpha$ benchmark test $\varphi_N = \1\{Q_N > c_{1-\alpha}\}$ for $H_0: \vT \vmu = \vnull$, where $c_{1-\alpha}$ is the $(1-\alpha)$-quantile of $Z$. However, the distribution of $Z$ depends on the unknown variances $\sigma^2_{is}, i=1, \dots, a,~s=1, \dots, d$ so that $\varphi_N$ is infeasible for most practical situations. For this reason we consider different bootstrap approaches in order to approximate the unknown limiting distribution and to derive 
adequate and asymptotically correct inference procedures based on $Q_N$  in \eqref{modATS}. This} will be explained in detail in the next section. 
Apart from statistical test decisions discussed in Section \ref{tests}, a central part of statistical analyses is the construction of confidence intervals, which allows for deeper insight into the variability and the magnitude of effects. This univariate concept can be generalized to multivariate endpoints by constructing multivariate confidence regions and simultaneous confidence intervals for contrasts $\vh^\top\vmu$ for any contrast vector $\vh \in \mathbb{R}^{ad}$ of interest. 
Details on the derivation of such confidence regions for $\vh^\top\vmu$ are given in Section \ref{confint} below.


\section{Bootstrap Procedures}\label{sec:bs}
The first bootstrap procedure we consider is a parametric bootstrap approach as proposed by \cite{Kon:2015} for the WTS. The second one is a wild bootstrap approach, which has already been successfully applied in the context of repeated measures or clustered data, see \cite{cameron2008bootstrap, cameron2015practitioner} or \cite{Friedrich2016}. {\color{black} The third procedure is a group-wise, nonparametric bootstrap approach.}
{\color{black}All of these} bootstrap approaches are based on the test statistic $Q_N$ in \eqref{modATS}. Note that the procedures derived in the following can also be used for multiple testing problems, either by applying the closed testing principle \citep{marcus1976closed, sonnemann2008general} or in the context of simultaneous contrast tests \citep{hasler2008multiple, hothorn2008simultaneous}.

\subsection{A Parametric Bootstrap Approach}
This asymptotic model based bootstrap approach has successfully been used 
in univariate one- and two-way factorial designs (\cite{krishnamoorthy2010, xu2013}), and has 
recently been applied to Wald-type statistics for general MANOVA by \cite{Kon:2015} and \cite{smaga2016bootstrap}, additionally assuming finite fourth moments. The approach is as follows: 
Given the data, we generate a parametric bootstrap sample as
\bqa
\vX_{i1}^*, \dots, \vX_{in_i}^* \stackrel{\text{i.i.d.}}{\sim} \mathcal{N}(\vnull, \vVh_i), \qquad i=1, \dots, a.
\eqa
The idea behind this method is to obtain an accurate finite sample approximation by mimicking the covariance structure given in the observed data. This is achieved by calculating $Q_N^*$ from the bootstrap variables $\vX_{i1}^*, \dots, \vX_{in_i}^*$, i.e.,
\bqan \label{eq: PBS stat}
Q_N^* = N (\volX^*_{\bullet})^\top \vT (\vT \vwhD^*_N \vT)^+\vT \volX^*_{\bullet}.
\eqan
We then obtain a parametric bootstrap test by comparing the original test statistic $Q_N$ with the conditional $(1-\alpha)$-quantile $c^*_{1-\alpha}$ of its bootstrap version $Q_N^*$.

\begin{theorem}\label{theo:MATS_pbs}
	The conditional distribution of $Q_N^*$ {\color{black}weakly approximates} the null distribution of $Q_N$ in probability for any parameter $\vmu \in \renu^{ad}$ and $\vmu_0$ with $\vT \vmu_0 = \vnull$, i.e.,
$$
\sup_{x \in \renu}|\Pr_{\vmu}(Q_N^*\leq x|\vX) - \Pr_{\vmu_0}(Q_N \leq x)| \stackrel{\Pr}{\to} 0,
$$
where $\Pr_{\vmu}(Q_N \leq x)$ and $\Pr_{\vmu}(Q_N^* \leq x | \vX)$ denote the unconditional and conditional distribution of $Q_N$ and $Q_N^*$, respectively, if $\vmu$ is the true underlying mean vector.
\end{theorem}

\subsection{A Wild Bootstrap Approach}
Another resampling approach, which is based on multiplying the fixed data with random weights, is the so-called wild bootstrap procedure. 
To this end, let $W_{ik}$ denote i.i.d.~random variables, independent of $\vX$, with $\E(W_{ik})=0, \Var(W_{ik})=1$ and $\sup_{i, k}\E(W_{ik}^4) < \infty$.
We obtain a bootstrap sample as
\bqa
\vX^\star_{ik} = W_{ik} (\vX_{ik}-\volX_{i \cdot}), i = 1, \dots, a, ~ k = 1, \dots, n_i.
\eqa
Note that there are different choices for the random weights $W_{ik}$, e.g., Rademacher distributed random variables \citep{davidson2008wild} or weights satisfying different moment conditions, see, e.g., \cite{ Beyersmann2013, liu1988bootstrap, mammen93, wu1986jackknife}. The choice of the weights typically depends on the situation. In our simulation studies, we have investigated the performance of different weights such as Rademacher distributed as well as $\mathcal{N}(0, 1)$ distributed weights (see \cite{lin1997non} for this specific choice). The results were rather similar and we therefore only present the results of our simulation study for standard normally distributed weights in Section \ref{sim} below.

Based on the bootstrap variables $\vX_{ik}^\star$, we can calculate $Q_N^\star$ in the same way as described 
for $Q_N^*$ in \eqref{eq: PBS stat} above. 
A wild bootstrap test is finally obtained by comparing $Q_N$ to the conditional $(1-\alpha)$-quantile of its wild bootstrap version $Q_N^\star$.

\begin{theorem}\label{theo:MATS_wild}
	The conditional distribution of $Q_N^\star$ {\color{black}weakly approximates} the null distribution of $Q_N$ in probability for any parameter $\vmu \in \renu^{ad}$ and $\vmu_0$ with $\vT \vmu_0 = \vnull$, i.e.,
$$
\sup_{x \in \renu}|\Pr_{\vmu}(Q_N^\star\leq x|\vX) - \Pr_{\vmu_0}(Q_N \leq x)| \stackrel{\Pr}{\to} 0.
$$
\end{theorem}

{\color{black}
\subsection{A nonparametric bootstrap approach}

The third approach we consider is the nonparametric bootstrap. Here, for each group $i=1, \dots, a$, we randomly draw $n_i$ independent selections $\vX^\dagger_{ik}$ with replacement from the $i$-th sample $\{\vX_{i1},\dots,\vX_{in_i}\}$. The bootstrap test statistic $Q_N^\dagger$ is then calculated in the same way as described above, i.e., by recalculating $Q_N$ with $\vX^\dagger_{ik}, i=1,\dots,a,\, k=1,\dots,n_i$. Finally, a nonparametric bootstrap test is obtained by comparing the original test statistic $Q_N$ to the empirical $(1-\alpha)$-quantile of $Q_N^\dagger$. The asymptotic validity of this method is guaranteed by

\begin{theorem}\label{theo:MATS_NPBS}
	The conditional distribution of $Q_N^\dagger$ weakly approximates the null distribution of $Q_N$ in probability for any parameter $\vmu \in \renu^{ad}$ and $\vmu_0$ with $\vT \vmu_0 = \vnull$, i.e.,
	$$
	\sup_{x \in \renu}|\Pr_{\vmu}(Q_N^\dagger \leq x|\vX) - \Pr_{\vmu_0}(Q_N \leq x)| \stackrel{\Pr}{\to} 0.
	$$
\end{theorem}

}

\section{Statistical Applications}\label{sec:stats}

We now want to base statistical inference on the modified test statistic in \eqref{modATS} using the bootstrap approaches described above. 
A thorough statistical analysis should ideally consist of two parts: First, statistical tests give insight into significant effects of the different factors as well as possible interactions. We therefore consider important properties of statistical tests based on the bootstrap approaches in Section \ref{tests}. Second, it is helpful to construct confidence regions for the unknown parameters of interest in order to gain a more detailed insight into the nature of the estimates. The derivation of such confidence regions is discussed in Section \ref{confint}.

\subsection{Statistical Tests}\label{tests}
In this section, we analyze the statistical properties of the bootstrap procedures described above. For ease of notation, we will only state the results for the parametric bootstrap procedure, i.e., consider the test statistic $Q_N^*$ based on $\vX_{ik}^*$ throughout. Note, however, that the results are also valid for the wild {\color{black}and the nonparametric} bootstrap procedure, i.e., the test statistics $Q_N^\star$ {\color{black} and $Q_N^\dagger$}.

As mentioned above, a bootstrap test $\varphi^*=\1\{Q_N > c^*_{1-\alpha}\}$ is obtained by comparing the original test statistic $Q_N$ to the $(1-\alpha)$-quantile $c_{1-\alpha}^*$ of its bootstrap version. In particular, $p$-values are numerically computed as follows:
\bit
\item[(1)] Given the data $\vX$, calculate the MATS $Q_N$ for the null hypothesis of interest.
\item[(2)] Bootstrap the data with either of the bootstrap approaches described above and calculate the corresponding test statistic $Q_N^{*,1}$.
\item[(3)] Repeat step (2) a large number of times, e.g., $B=10,000$ times, and obtain values $Q_N^{*,1}, \dots, Q_N^{*, B}$.
\item[(4)] Calculate the $p$-value based on the empirical distribution of $Q_N^{*,1}, \dots, Q_N^{*, B}$ as
$$
\textrm{p-value} = \frac{1}{B}\sum_{b=1}^B \1\{Q_N \leq Q_N^{*, b}\}.
$$
\eit
Theorems \ref{theo:MATS_pbs} -- \ref{theo:MATS_NPBS} imply that the corresponding tests asymptotically keep the pre-assigned level $\alpha$ under the null hypothesis and are consistent for any fixed alternative $\vT \vmu \neq \vnull$, i.e., $E_{\vmu}(\varphi^*) \to \alpha \cdot \1\{\vT\vmu=\vnull\}+\1\{\vT\vmu \neq \vnull\}$.
Moreover, for local alternatives $H_1: \vT \vmu = \sqrt{N}^{-1}\vT \vnu,  \vnu \in \mathbb{R}^{ad}$, the bootstrap tests have the same asymptotic power as $\varphi_N = \1\{Q_N > c_{1-\alpha}\}$, where $c_{1-\alpha}$ is the $(1-\alpha)$-quantile of $Z$ given in Theorem \ref{theo:MATS}.
In particular, the asymptotic relative efficiency of the bootstrap tests compared to $\varphi_N$ is 1 in this situation.

\subsection{Confidence regions and confidence intervals for contrasts}\label{confint}

In order to conduct a thorough statistical analysis, interpretation of the results should not be based on $p$-values alone. In addition, it is helpful to construct confidence regions for the unknown parameter. The concept of a confidence region is the same as that of a confidence interval in the univariate setting: We want to construct a multivariate region, which is likely to contain the true, but unknown parameter of interest.
The aim of this section is to derive multivariate confidence regions and simultaneous confidence intervals for 
contrasts $\vh^\top\vmu$ for any contrast vector $\vh$ of interest.
Such contrasts include, e.g., the difference in means $\vmu_1 - \vmu_2$ in two-sample problems, Dunnett's many-to-one comparisons, Tukey's all-pairwise comparisons, and many more, see, e.g., \cite{hothorn2008simultaneous} for specific examples. 
In this section, we will base the derivation of confidence regions on the bootstrap approximations given in Section \ref{sec:bs}, i.e., we will use one of the bootstrap quantiles. Again, we only formulate the results for $c^*_{1-\alpha}$.

{\color{black}
For the derivation of a confidence region, first note that the results
from Section \ref{tests} {\color{black}imply} that the null hypothesis {\color{black} $H_0:\vH\vmu=\vH\vmu_0$ for a vector of contrasts $\vH\vmu_0, \vH = (\vh_1| \dots| \vh_q)^\top \in \mathbb{R}^{q \times ad}, \vmu_0\in \mathbb{R}^{ad},$ is
	rejected at asymptotic level $\alpha$, if 
	$
	N (\vH \volX_{\bullet}- \vH \vmu_0)^\top (\vH \vwhD_N\vH^\top)^+(\vH \volX_{\bullet} - \vH \vmu_0) 
	$ 
	is larger than the bootstrap quantile $c^*_{1-\alpha}$.
Thus, a	confidence region for {\color{black} the vector of} contrasts $\vH\vmu$ is determined by the set of all $\vH \vmu$ such that
	\bqa
	 N (\vH \volX_{\bullet}- \vH \vmu)^\top (\vH \vwhD_N\vH^\top)^+(\vH \volX_{\bullet} - \vH \vmu) \leq c_{1-\alpha}^*.
	\eqa	
	A confidence ellipsoid is now obtained based on the eigenvalues $\widehat{\lambda}_s$ and eigenvectors $\vwhe_s$ of $\vH \vwhD_N \vH^\top$. As in \cite{johnson}, the direction and lengths of {\color{black} its} axes 
	are determined by going $\sqrt{\widehat{\lambda}_s \cdot c_{1-\alpha}^*/N}$ units along the eigenvectors $\vwhe_s$ of $\vH \vwhD_N \vH^\top$. In other words, the axes of the ellipsoid are {\color{black} given by}
	\bqan \label{confreg}
	\vH \volX_{\bullet} \pm \sqrt{\widehat{\lambda}_s \cdot c_{1-\alpha}^*/N}\cdot  \vwhe_s, \qquad s=1, \dots, d.
	\eqan
	Note that this approach is similar to the construction of confidence intervals in the univariate case, where we exploit the one-to-one relationship between CIs and tests. 	
	While we can calculate \eqref{confreg} for arbitrary dimensions $d$, we cannot display the joint confidence region graphically for $d \geq 4$. 
	In the two-sample case with $d=2$ endpoints, however, the ellipse can be plotted: Beginning at the center $\vH\volX_{\bullet}$ the axes of the ellipsoid are given by $\pm \sqrt{\widehat{\lambda}_s \cdot c_{1-\alpha}^*/N}\cdot  \vwhe_s, ~ s= 1,2 $. That is, the confidence ellipse extends $\sqrt{\widehat{\lambda}_s \cdot c_{1-\alpha}^*/N}$ units along the estimated eigenvector $\vwhe_s$ for $s=1, 2$. Therefore, we get a graphical representation of the relation between the group-mean differences $\mu_{11}-\mu_{21}$ and $\mu_{12}-\mu_{22}$ of the first and second component, see Section \ref{datex2} and Figure \ref{confellipse} {\color{black}in the supplementary material} for an example.
	
}

{\color{black} Concerning the derivation of multiple contrast tests and simultaneous confidence intervals for contrasts, we consider the family of hypotheses
	$$
	\vOmega = \{ H_0: \vh^\top_\ell \vmu = \vnull \text{ with } \vh_\ell \neq \vnull, \ell =1, \dots, q\}.
	$$

As shown in Sections~\ref{mod} and \ref{sec:bs} a test statistic for testing the null hypothesis {\color{black} $H_0:\vH\vmu=\vnull$ is
given by $Q_N$ in \eqref{modATS}. 
 Consequently, working with a single contrast $\vh_\ell$ as contrast matrix leads to the test statistic 
$$
Q_N^\ell = N (\vh^\top_\ell \volX_{\bullet})^\top (\vh^\top_\ell \vwhD_N\vh_\ell)^{-1}(\vh^\top_\ell \volX_{\bullet})  = N \frac{\left(\sum_{i=1}^a \sum_{s=1}^d h_{\ell, is} \olX_{i \cdot s}\right)^2}{\sum_{i=1}^a \sum_{s=1}^d h_{\ell, is}^2 \widehat{\sigma}_{is}^2}
$$
for the null hypotheses $H_0^\ell: \vh^\top_\ell \vmu = \vnull, \ell =1, \dots, q$. 
  Here, $\vh_\ell=(h_{\ell, 11}, \dots, h_{\ell, ad})^\top$.
}
To obtain a single critical value with one of the bootstrap methods we may, e.g., consider the usual maximum or sum statistics. 
We exemplify the idea for the latter. Thus, let
$$
S_N \equiv N (\vH\volX_\bullet)^\top \diag \left(\left(\vh^\top_\ell \vwhD_N\vh_\ell
\right)^{-1}: \ell = 1, \dots, q \right) \vH \volX_\bullet = \sum_{\ell=1}^q Q_N^\ell
$$
and denote by ${q}_{1-\alpha}^*$ the conditional $(1-\alpha)$-quantile of its corresponding bootstrap version $S_N^*$.
From the proofs of Theorem \ref{theo:MATS_pbs} -- \ref{theo:MATS_NPBS} given in the supplement it follows for any of the three bootstrap methods described in Section~\ref{sec:bs} that 
 $\widehat{\sigma}_{is}^*$ are consistent estimates for ${\sigma}_{is}$ ($i=1,\dots,a, \, s=1,\dots,d$) and that 
 $ \sqrt{N} \vH\volX_\bullet^*$ asymptotically mimics the distribution of $ \sqrt{N} \vH(\volX_\bullet - \vmu)$. Thus, the continuous mapping theorem implies $\Pr\left(S_N \leq {q}^*_{1-\alpha}\right) \to 1-\alpha$ as $N \to \infty$ and therefore
$$
\Pr \left(\bigcap_{\ell = 1}^q \{Q_N^\ell \leq {q}^*_{1-\alpha}\}\right) \leq \Pr \left(\sum_{\ell=1}^q Q_N^\ell \leq {q}^*_{1-\alpha}\right) \to 1-\alpha, N \to \infty.
$$
This implies, that 
 simultaneous $100(1-\alpha)\%$ confidence interval{\color{black}s} for contrast{\color{black}s} $\vh_\ell^\top\vmu, {\color{black}\ell = 1, \dots, q,}$ {\color{black} are} given by
 \bqa
 \vh_\ell^\top\volX_{\bullet} \pm \sqrt{{q}^*_{1-\alpha} \cdot \vh_\ell^\top\vwhD_N\vh_\ell /N}.
 \eqa 

In the supplement we additionally explain that the bootstrap idea also works for the usual maximum statistic.
}}

\section{Simulations} \label{sim}
The procedures described in Section \ref{sec:bs} are valid for large sample sizes. In order to investigate their behavior for small samples, we have conducted various simulations. In the simulation studies, the behavior of the proposed approaches was compared to a parametric bootstrap approach for the WTS as in \cite{Kon:2015} since this turned out to perform better than other resampling versions of the WTS and Wilk's $\Lambda$. {\color{black} For comparison, we also included the asymptotic $chi^2$ approximation of the WTS.}
All simulations were conducted using \textsf{R} Version 3.3.1 \citep{R} each with nsim = 5,000 simulation and nboot = 5,000 bootstrap runs.
We investigated a one- and a two-factorial design. 

\subsection{One-way layout}\label{sim:oneway}
For the one-way layout, data was generated as in \cite{Kon:2015}. We considered $a=2$ treatment groups and $d \in \{4, 8\}$ endpoints as well as the following covariance settings:
\bqa
\text{Setting 1: }&& \vV_1 = \vI_d + 0.5 (\vJ_d-\vI_d) = \vV_2,\\
\text{Setting 2: }&& \vV_1 = \left((0.6)^{|r-s|}\right)_{r, s =1}^d = \vV_2,\\
\text{Setting 3: }&& \vV_1 = \vI_d + 0.5 (\vJ_d-\vI_d) \text{ and } \vV_2=\vI_d\cdot 3 + 0.5 (\vJ_d-\vI_d),\\
\text{Setting 4: }&& \vV_1 = \left((0.6)^{|r-s|}\right)_{r, s =1}^d \text{ and } \vV_2=\left((0.6)^{|r-s|}\right)_{r, s =1}^d + \vI_d \cdot 2.\\
\eqa
Setting 1 represents a compound symmetry structure, while setting 2 is an autoregressive covariance structure. Both settings 1 and 2 represent homoscedastic scenarios while settings 3 and 4 display two scenarios with unequal covariance structures.
Data was generated by
\bqa
\vX_{ik} = \vmu_i + \vV_i^{1/2} \vep_{ik}, ~i=1, \dots, a;~ k=1, \dots, n_i,
\eqa
where $\vV_i^{1/2}$ denotes the square root of the matrix $\vV_i$, i.e., $\vV_i =\vV_i^{1/2} \cdot \vV_i^{1/2}$. The mean vectors $\vmu_i$ were set to $\vnull$ in both groups.
The i.i.d.~random errors $\vep_{ik}=(\ep_{ik1}, \dots, \ep_{ikd})^\top$ with mean $\E(\vep_{ik})= \vnull_d$ and $\Cov(\vep_{ik}) = \vI_{d \times d}$ were generated by simulating independent standardized components
\bqa
\epsilon_{iks}= \frac{Y_{iks}- \E(Y_{iks})}{\sqrt{\Var(Y_{iks})}}
\eqa
for various distributions of $Y_{iks}$. In particular, we simulated normal, $\chi^2_3$, lognormal, $t_3$ and double-exponential distributed random variables.
We investigated balanced as well as unbalanced designs with sample size vectors $\vn^{(1)}=(10, 10)^\top$, $\vn^{(2)} = (20, 20)^\top$, $\vn^{(3)} = (10, 20)^\top$ and $\vn^{(4)}=(20, 10)^\top$, respectively. A major criterion concerning the accuracy of the procedures is their behavior in situations where increasing variances (settings 3 and 4 above) are combined with increasing sample sizes ($\vn^{(3)}$, positive pairing) or decreasing sample sizes ($\vn^{(4)}$, negative pairing).\\
In this setting, we tested the null hypothesis $H_0^{\mu}: \{(\vP_a \kp \vI_d)\vmu\ = \vnull \}=\{\vmu_1 = \vmu_2\}$, i.e., no treatment effect.
The resulting type-I error rates (nominal level $\alpha = 5\% $) for $d=4$ and $d=8$ endpoints are displayed in Table~\ref{tab1} (normal distribution) and Table~\ref{tab2} ($\chi^2_3$-distribution), respectively. Further simulation results for lognormal, $t_3$ and double-exponential distributed errors {\color{black} for the parametric bootstrap of WTS and MATS} can be found in Tables \ref{app:logn} -- \ref{app:dexp} in the supplementary material.

{\color{black}As already noticed by \cite{Kon:2015}, the WTS with the $\chi^2$ approximation is far too liberal, reaching type-I error rates of more than 50\% in some scenarios (e.g., for $d=8$ with negative pairing, i.e., covariance setting S3 and $\vn=(20, 10)^\top$). Even in the scenarios with only $d=4$ dimensions and $\vn = (20, 20)^\top$, the error rates are around 9\% instead of 5\%. The parametric bootstrap of the WTS greatly improves this behavior for all situations. However, it still shows a rather liberal behavior with type-I error rates of around 10\% in some situations, e.g., $d=8$ dimensions with S3 or S4 and $\vn=(20, 10)^\top$ in Tables~\ref{tab1} and \ref{tab2}.
  
The wild bootstrap of the MATS shows a rather liberal behavior across all scenarios and can therefore not be recommended in practice. In contrast, both the parametric and the nonparametric bootstrap of the MATS show a very accurate type-I error rate control. The nonparametric bootstrap is often slightly more conservative than the parametric bootstrap and thus works better in situations with negative pairing, especially for the $\chi^2_3$-distribution, i.e., for S3 and S4 with $\vn=(20, 10)^\top$ and $d=4$ or $d=8$ dimensions in Table \ref{tab2}. In most other scenarios, however, the parametric bootstrap yields slightly better results. The improvement of the parametric bootstrap MATS over WTS (PBS) and nonparametric bootstrap MATS is most pronounced for large $d$, i.e., in situations where $d$ is close to $\min(n_1, n_2)$.
  
  }

However, in situations with negative pairing and skewed distributions (see Table \ref{tab2} as well as Table \ref{app:logn} in the supplementary material), the parametric bootstrap MATS shows a slightly liberal behavior. For $t_3$ and double-exponentially distributed errors and negative pairing, in contrast, the parametric bootstrap MATS is slightly conservative, see Tables \ref{app:t3} and \ref{app:dexp} in the supplementary material, respectively.\\
Surprisingly, {\color{black} the resampling approaches based on the MATS} improve with growing $d$ in most settings, i.e., when the number of endpoints is closer to the sample size. The WTS approach, in contrast, gets worse in these scenarios. This might be an interesting approach for future research in high-dimensional settings such as in \cite{PEB}. 

As a result, we find that the MATS with the parametric bootstrap approximation is the best procedure in most scenarios. Especially, it is less conservative than the {\color{black} nonparametric} bootstrap approximation and less liberal than the WTS equipped with the parametric bootstrap approach over all simulation settings. Only in situations with negative pairing and skewed distributions, the new procedure shows a slightly liberal behavior. 

\begin{table}[H]
	\centering
	\caption{Type-I error rates in \% (nominal level $\alpha = 5\%$) for the {\color{black} WTS with $\chi^2$-approximation and parametric bootstrap (PBS) and the MATS with wild bootstrap (wild), parametric bootstrap (PBS) and nonparametric bootstrap (NPBS) in the one-way layout for the normal distribution.}}
	\label{tab1}
	\begin{tabular}{c|c|c|cc|ccc}
		\hline
		$d$ & Cov & $\vn$ & WTS ($\chi^2$) & WTS (PBS) & MATS (wild) & MATS (PBS) & MATS (NPBS) \\ 
		\hline
		\multirow{16}{*}{$d=4$} & \multirow{4}{*}{S1}   & (10, 10) & 15.2 & 4.5 & 6.9 & 5.2 & 4.4 \\ 
		&  & (10, 20) & 14.5 & 5.9 & 6.9 & 5.1 & 4.5 \\ 
		&  & (20, 10) & 14 & 5.6 & 7.2 & 5.3 & 4.8 \\ 
		&  & (20, 20) & 9.5 & 5.3 & 6 & 5.1 & 5.1 \\ \cline{2-8}
		& \multirow{4}{*}{S2} & (10, 10) & 15.2 & 4.5 & 7 & 5 & 4.5 \\ 
		&  & (10, 20) & 14.5 & 5.8 & 6.9 & 5 & 4.5 \\ 
		&  & (20, 10) & 14 & 5.6 & 7.3 & 5.5 & 4.9 \\ 
		&  & (20, 20) &   9.5 & 5.4 & 6.3 & 5.2 & 5 \\  \cline{2-8}
		& \multirow{4}{*}{S3} & (10, 10) & 18.3 & 5.5 & 7.3 & 4.8 & 3.6 \\ 
		&  & (10, 20) & 10.9 & 4.7 & 6.4 & 4.8 & 4.4 \\ 
		&  & (20, 10) & 21.4 & 6.6 & 7.8 & 4.8 & 3.4 \\ 
		&  & (20, 20) &  11.2 & 5.7 & 6.3 & 5.1 & 4.6 \\  \cline{2-8}
		& \multirow{4}{*}{S4} & (10, 10) & 18.3 & 5.6 & 7.5 & 4.8 & 3.9 \\ 
		&  & (10, 20) & 11 & 5.2 & 6.1 & 4.6 & 4.3 \\ 
		&  & (20, 10) &21 & 6.7 & 7.9 & 4.7 & 3.2 \\ 
		&  & (20, 20) &     10.9      &   5.7      &   6.2    &     5.0       &  4.7  \\ 
		\hline \hline
		\multirow{16}{*}{$d=8$} & \multirow{4}{*}{S1} & (10, 10) & 38.6 & 4.7 & 7.7 & 5.1 & 4.3 \\ 
		&  & (10, 20) & 31 & 6.2 & 6.9 & 5 & 4.2 \\ 
		&  & (20, 10) & 32.1 & 6.1 & 6.6 & 4.6 & 4 \\ 
		&  & (20, 20) &  17.0      &   4.9      &   5.8      &   4.8       &  4.8   \\ \cline{2-8}
		& \multirow{4}{*}{S2} & (10, 10) & 38.6 & 4.5 & 7.9 & 4.3 & 3.4 \\ 
		&  & (10, 20) & 31 & 6.3 & 7.4 & 4.3 & 3.6 \\ 
		&  & (20, 10) & 32.1 & 6.1 & 7 & 4.1 & 3.4 \\ 
		&  & (20, 20) & 17.0    &     4.7      &   6.2      &   4.8      &   4.5  \\ \cline{2-8}
		& \multirow{4}{*}{S3} & (10, 10) & 50.1 & 6.6 & 7.9 & 4.2 & 2.8 \\ 
		&  & (10, 20) & 21.8 & 4.1 & 6.3 & 4.4 & 4.1 \\ 
		&  & (20, 10) & 55 & 10.3 & 8.5 & 3.6 & 2.2 \\ 
		&  & (20, 20) &  21.9     &    5.4    &     6.1    &     4.0     &    3.6  \\ \cline{2-8}
		& \multirow{4}{*}{S4} & (10, 10) & 48.9 & 6.3 & 7.8 & 3.6 & 2.4 \\ 
		&  & (10, 20) & 21.9 & 4.2 & 6.3 & 3.8 & 3.4 \\ 
		&  & (20, 10) &  54.1 & 10.4 & 8.4 & 3.5 & 2 \\  
		&  & (20, 20) & 21.8     &    5.2    &     6.0      &   3.9      &   3.6  \\ 
		\hline
	\end{tabular}
\end{table}

\begin{table}[H]
	\centering
	\caption{Type-I error rates in \% (nominal level $\alpha = 5\%$) for the {\color{black} WTS with $\chi^2$-approximation and parametric bootstrap (PBS) and the MATS with wild bootstrap (wild), parametric bootstrap (PBS) and nonparametric bootstrap (NPBS) in the one-way layout for the $\chi^2_3$-distribution.}}
	\label{tab2}
	\begin{tabular}{c|c|c|cc|ccc}
		\hline
		$d$ & Cov & $\vn$ & WTS ($\chi^2$) & WTS (PBS) & MATS (wild) & MATS (PBS) & MATS (NPBS) \\ 
		\hline
		\multirow{16}{*}{$d=4$} & \multirow{4}{*}{S1} & (10, 10) &  15.3 & 4 & 7.1 & 4.8 & 3.4 \\ 
		&  & (10, 20) & 13.9 & 5.5 & 7.3 & 5.6 & 4.6 \\ 
		&  & (20, 10) & 14.6 & 5.7 & 7.7 & 5.9 & 4.6 \\ 
		&  & (20, 20) & 8.9 & 4.7 & 6.3 & 5.5 & 5 \\ \cline{2-8}
		& \multirow{4}{*}{S2} & (10, 10) & 15.3 & 4.1 & 7.1 & 4.5 & 3.2 \\ 
		&  & (10, 20) & 13.9 & 5.6 & 7.5 & 5.5 & 4.5 \\ 
		&  & (20, 10) & 14.6 & 5.8 & 7.7 & 5.5 & 4.5 \\ 
		&  & (20, 20) & 8.9 & 4.7 & 6.3 & 5.3 & 4.7 \\ \cline{2-8}
		& \multirow{4}{*}{S3} & (10, 10) & 20.6 & 7.1 & 9.5 & 6.1 & 3.8 \\ 
		&  & (10, 20) & 11.2 & 4.8 & 6.9 & 4.8 & 3.7 \\ 
		&  & (20, 10) & 26.2 & 10.9 & 12.3 & 8.9 & 5.6 \\ 
		&  & (20, 20) & 12.8 & 6.6 & 7.6 & 6 & 4.7 \\ \cline{2-8}
		& \multirow{4}{*}{S4} & (10, 10) & 21.2 & 7.2 & 9.6 & 6.2 & 3.8 \\ 
		&  & (10, 20) & 11.1 & 5 & 6.9 & 4.7 & 3.4 \\ 
		&  & (20, 10) & 26.5 & 10.7 & 12.7 & 8.9 & 5.6 \\ 
		&  & (20, 20) &  12.9     &    6.7     &    7.7    &     6.2      &   4.7   \\ 
		\hline \hline
		\multirow{16}{*}{$d=8$} & \multirow{4}{*}{S1} & (10, 10)  & 39.3 & 3.8 & 7.7 & 4.9 & 3.4 \\ 
		&  & (10, 20) & 32.3 & 5.5 & 7.6 & 5.9 & 4.7 \\ 
		&  & (20, 10) & 33.4 & 6.3 & 7.2 & 5.1 & 4.2 \\ 
		&  & (20, 20) & 16.9 & 4.5 & 5.9 & 4.9 & 4.6 \\ \cline{2-8}
		& \multirow{4}{*}{S2} & (10, 10) & 39.3 & 3.8 & 8.1 & 4.3 & 2.7 \\ 
		&  & (10, 20) & 32.3 & 5.5 & 8.6 & 5.2 & 4 \\ 
		&  & (20, 10) & 33.4 & 6.3 & 7.6 & 4.9 & 3.9 \\ 
		&  & (20, 20) &  16.9    &     4.5      &   6.2  &      4.5      &   4.0   \\ \cline{2-8}
		& \multirow{4}{*}{S3} & (10, 10) & 53.1 & 6.8 & 10.2 & 5.5 & 3.1 \\ 
		&  & (10, 20) & 23.4 & 4.8 & 6.8 & 4.6 & 3.5 \\ 
		&  & (20, 10) & 59.9 & 13.9 & 13.7 & 8.1 & 4.6 \\ 
		&  & (20, 20) &  24.8    &     6.9      &   7.9   &      5.7     &    4.1   \\ \cline{2-8}
		& \multirow{4}{*}{S4} & (10, 10) & 52.5 & 6.3 & 11 & 5.3 & 2.6 \\ 
		&  & (10, 20) & 24.3 & 4.5 & 7.1 & 4.1 & 2.8 \\ 
		&  & (20, 10) & 59 & 13.6 & 14.8 & 8.4 & 4.5 \\ 
		&  & (20, 20) &    24.3     &    6.9     &    7.7      &   5.7     &    4.0   \\ 
		\hline
	\end{tabular}
\end{table}

\subsubsection{Singular Covariance Matrix}\label{sim:degCovmat}

In order to analyze the behavior of the discussed methods in designs involving singular covariance matrices, we considered the one-way layout described above with $a=2$ groups and $d\in \{4, 8\}$ observations involving the following covariance settings (displayed for $d=4$):
\bqa
\text{Setting 5: }&& \vV_1 = \left (\begin{array}{cccc} 1 & 1/2 & 1 & 1\\
	1/2 &  1 &  1/2 &  1/2\\
	1 &  1/2 & 1 &  1 \\
	1&  1/2 & 1&  1 \\
\end{array} \right ), \vV_2 = \vV_1 + 0.5\cdot \vJ_d\\
\text{Setting 6: }&& \vV_1 = \left (\begin{array}{cccc} 1 & 0.6 & 0.36 & 0.18\\
	0.6 &  1 &  0.6 &  0.3\\
	0.36 &  0.6 & 1 &  0.5 \\
	0.18&  0.3 & 0.5 &  0.25 \\
\end{array} \right ), \vV_2 = \vV_1 + 0.5\cdot \vJ_d\\
\text{Setting 7: }&& \vV_1 = \left (\begin{array}{cccc}  1 & 0 &  0  & 0\\
	0 & \sqrt{2} &   0 & 0\\
	0 &  0 & 2 &  1 \\
	0&  0 & 1&  0.5 \\
\end{array} \right ), \vV_2 = \vV_1 + 0.5\cdot \vJ_d\\
\eqa

Setting 6 is based on an $AR(0.6)$ covariance matrix (see setting 2 above), where the last row and column have been replaced by half the row/column before, respectively. Setting 7 is based on $\vwtV_1=\diag(\sqrt{2^s}), s= 0, \dots, d-1$, where the last row and column have been replaced by 
half the row/column before.
We have considered {\color{black} the same sample size vectors as above.}

The results are displayed in Tables \ref{deg:normal} and \ref{deg:chil}.
The parametric bootstrap of the MATS again yields the best results {\color{black} in almost all scenarios. The wild bootstrap, in contrast, is again rather liberal. 
For the $\chi^2$ approximation of the WTS, the results are in concordance with the theoretical reflections mentioned in Section \ref{mod}: Covariance setting S5 corresponds to the case, where the rank of $\vT$ and $\vT \vSigma \vT$ differs and as calculated above, the $\chi^2$-approximation becomes very conservative here. In setting S6 and S7, in contrast, there is no rank jump despite the singular covariance matrices and the $\chi^2$-approximation shows its usual liberal behavior. 
Since the rank of $\vT \vSigma\vT$ is not known in practice, the WTS should not be used for data with possibly singular covariance matrices.
 It turns out, however, that the parametric bootstrap of the WTS is relatively robust against singular covariance matrices. Its behavior is comparable to the scenarios above with non-singular covariance matrices. It is, however, rather liberal for $\vn=(20, 10)^\top$, especially with the $\chi^2_3$-distribution, see Table \ref{deg:chil}. This behavior is improved by the parametric bootstrap MATS, e.g., for $d=8$ and S7, the WTS (PBS) leads to a type-I error of 9\%, whereas the MATS (PBS) is at 5.1\%. The nonparametric bootstrap, in contrast,  sometimes leads to strictly conservative test decisions. This is especially apparent for $d=8$ and covariance setting S7 in Tables \ref{deg:normal} and \ref{deg:chil}.
}

\begin{table}[h]
	\centering
	\caption{Type-I error rates in \% (nominal level $\alpha = 5\%$) for the {\color{black} WTS with $\chi^2$-approximation and parametric bootstrap (PBS) and the MATS with wild bootstrap (wild), parametric bootstrap (PBS) and nonparametric bootstrap (NPBS) in the one-way layout with singular covariance matrices for the normal distribution.}}
	\label{deg:normal}
	\begin{tabular}{c|c|c|cc|ccc}
		\hline
		$d$ & Cov & $\vn$ & WTS ($\chi^2$) & WTS (PBS) &  MATS (wild) & MATS (PBS) & MATS (NPBS)\\ 
		\hline
		\multirow{12}{*}{$d=4$} & \multirow{4}{*}{S5} & (10, 10)  & 3.1 & 5.1 & 5.9 & 4.8 & 4.4 \\ 
		&  & (10, 20) & 2.6 & 5.1 & 5.7 & 4.9 & 4.6 \\ 
		&  & (20, 10) & 2.7 & 5.3 & 6.6 & 5.3 & 4.7 \\ 
		&  & (20, 20) &  1.7    &     4.8    &     5.6   &      5.2    &     5.0   \\ \cline{2-8}
		& \multirow{4}{*}{S6} & (10, 10) & 16.7 & 5.3 & 7.1 & 5 & 4.6 \\ 
		&  & (10, 20) & 12.4 & 5.1 & 6 & 4.7 & 4.3 \\ 
		&  & (20, 10) & 17.1 & 6.1 & 7.1 & 5.3 & 4.6 \\ 
		&  & (20, 20) & 10.2      &   5.5     &    6.0     &    5.2     &    5.1  \\ \cline{2-8}
		& \multirow{4}{*}{S7} & (10, 10) & 16.6 & 5.2 & 7.3 & 4.7 & 4 \\ 
		&  & (10, 20) & 12.3 & 5.8 & 6.6 & 4.6 & 4.2 \\ 
		&  & (20, 10) & 16.3 & 5.7 & 6.9 & 4.5 & 4 \\ 
		&  & (20, 20) &  9.4    &     4.8     &    5.9    &     4.8     &    4.8   \\ 
		\hline\hline
		\multirow{12}{*}{$d=8$} & \multirow{4}{*}{S5} & (10, 10) & 2.8 & 4.5 & 6.2 & 5 & 4.7 \\ 
		&  & (10, 20) & 2.3 & 4.9 & 5.5 & 4.9 & 4.7 \\ 
		&  & (20, 10) & 2.6 & 4.4 & 5.6 & 4.7 & 4.3 \\ 
		&  & (20, 20) &   1.5    &     4.6    &     5.5  &      4.9     &    4.8   \\ \cline{2-8}
		& \multirow{4}{*}{S6} & (10, 10) & 39.5 & 4.4 & 8.2 & 5 & 4.2 \\ 
		&  & (10, 20) & 28.8 & 5.4 & 6.6 & 4.6 & 4.2 \\ 
		&  & (20, 10) &   35.7   &      7.0    &     7.5   &      4.8      &   4.0   \\ 
		&  & (20, 20) &   17.3    &     4.7    &     6.1   &    4.8     &    4.5   \\ \cline{2-8}
		& \multirow{4}{*}{S7} & (10, 10) & 38.8 & 4.2 & 7.4 & 4 & 2.9 \\ 
		&  & (10, 20) & 27.4 & 5.2 & 6.6 & 3.6 & 3 \\ 
		&  & (20, 10) & 36.3 & 6.3 & 7.4 & 3.8 & 3.2 \\ 
		&  & (20, 20) &  17.3    &     5.1  &     5.9   &      4.2      &   4.0   \\ 
		\hline
	\end{tabular}
\end{table}

\begin{table}[H]
	\centering
	\caption{Type-I error rates in \% (nominal level $\alpha = 5\%$) for the {\color{black} WTS with $\chi^2$-approximation and parametric bootstrap (PBS) and the MATS with wild bootstrap (wild), parametric bootstrap (PBS) and nonparametric bootstrap (NPBS) in the one-way layout with singular covariance matrices for the $\chi^2_3$-distribution.}}
	\label{deg:chil}
	\begin{tabular}{c|c|c|cc|ccc}
		\hline
		$d$ & Cov & $\vn$ & WTS ($\chi^2$) & WTS (PBS) &  MATS (wild) & MATS (PBS) & MATS (NPBS)\\ 
		\hline
		\multirow{12}{*}{$d=4$} & \multirow{4}{*}{S5} & (10, 10) & 2.7 & 4.2 & 6.7 & 5.4 & 4.5 \\ 
		&  & (10, 20) & 1.9 & 4.8 & 5.9 & 4.9 & 4.5 \\ 
		&  & (20, 10) & 3.5 & 6.5 & 7.3 & 5.9 & 5.2 \\ 
		&  & (20, 20) &    1.7     &    4.9     &    6.2     &    5.7       &  5.5   \\ \cline{2-8}
		& \multirow{4}{*}{S6} & (10, 10) & 19.7 & 7.1 & 7.4 & 5.2 & 4.1 \\ 
		&  & (10, 20) & 14.6 & 7.1 & 6.4 & 5 & 4.3 \\ 
		&  & (20, 10) & 20.1 & 8.5 & 8.1 & 6.4 & 5.4 \\ 
		&  & (20, 20) &   11.4    &    6.3     &    6.5     &    5.7     &    5.3   \\ \cline{2-8}
		& \multirow{4}{*}{S7} & (10, 10) & 19.4 & 7 & 7.1 & 4.1 & 3.1 \\ 
		&  & (10, 20) & 14.5 & 6.7 & 6.4 & 4.2 & 3.4 \\ 
		&  & (20, 10) & 20.3 & 8.7 & 8.3 & 6.1 & 4.5 \\ 
		&  & (20, 20) &    11.7      &   6.4     &    6.1   &     5.1      &   4.5   \\ 
		\hline\hline
		\multirow{12}{*}{$d=8$} & \multirow{4}{*}{S5} & (10, 10)  & 2.4 & 4.7 & 6.1 & 5.1 & 4.6 \\ 
		&  & (10, 20) & 2.6 & 5.3 & 6.1 & 5.3 & 5 \\ 
		&  & (20, 10) & 3 & 5.6 & 6 & 5.1 & 4.6 \\ 
		&  & (20, 20) & 1.2      &   4.5   &      5.9    &     5.3     &    5.1 \\ \cline{2-8}
		& \multirow{4}{*}{S6} & (10, 10) & 43.1 & 5.4 & 8.2 & 5.1 & 3.9 \\ 
		&  & (10, 20) & 30.7 & 6.6 & 7.3 & 5.2 & 4.2 \\ 
		&  & (20, 10) & 39.2     &   8.7   &     8.3   &      5.6      &   4.5   \\ 
		&  & (20, 20) &    19.3    &     5.6     &    6.8     &    5.1     &    4.7   \\ \cline{2-8}
		& \multirow{4}{*}{S7} & (10, 10) & 42.4 & 5.5 & 7.5 & 3.3 & 1.7 \\ 
		&  & (10, 20) & 31.1 & 6.3 & 7.1 & 4 & 2.4 \\ 
		&  & (20, 10) & 39.5 & 9 & 9.2 & 5.1 & 3.4 \\ 
		&  & (20, 20) & 18.7     &    5.3     &    5.1    &     3.2      &   2.6  \\ 
		\hline
	\end{tabular}
\end{table}

%

\subsection{Two-way layout}
{\color{black} We have investigated the behavior of the methods in a setting with two crossed factors $A$ and $B$, which is again adapted from \cite{Kon:2015}. In particular, we simulated a $2\times2$ designs with covariance matrices similar to the one-way layout above. A detailed description of the simulation settings as well as the results for the main and interaction effects are deferred to the supplementary material. Here we only summarize our findings:} 
Since the total sample size $N$ is larger in this scenario, the asymptotic results come into play and therefore all methods lead to more accurate results than in the one-way layout. {\color{black}Nevertheless we} find a similar behavior as in the one-way layout: Again, the MATS and the WTS with the parametric bootstrap approach control the type-I error very accurately, whereas the {\color{black} nonparametric bootstrap approach leads to slightly more conservative results. Both the WTS with $\chi^2$ approximation and the wild bootstrap MATS can not be recommended due to their liberal behavior.
In situations with negative pairing (covariance setting 10 and 11 with sample size vector $\vn^{(3)}$), the parametric bootstrap MATS improves the slightly liberal behavior of the WTS, see e.g., Table \ref{app:facA} for the normal distribution, where the WTS (PBS) leads to a type-I error of 6.1\%, while the MATS (PBS) is at 4.9\%.
}

{\color{black}
	\subsection{Power}
	
	We have investigated the empirical power of the proposed methods to detect a fixed alternative in the simulation scenarios above. Data was simulated as described
	in Section~\ref{sim:oneway} but now with $\vmu_1 =\vnull$ and $\vmu_2 = (\delta, \dots, \delta)^\top$ for varying shifts $\delta \in \{ 0, 0.5,  1,  1.5,  2,  3\}$. Due to the liberality of the classical Wald-type test and the wild bootstrapped MATS, we have only considered the WTS with parametric bootstrap as well as the parametric and nonparametric bootstrap of the MATS. The results for selected scenarios are displayed in Figures \ref{power_chi} -- \ref{power_n}. The plots show that both resampling versions of the MATS have a higher power for detecting the fixed alternative than the WTS. The parametric bootstrap of the MATS has a slightly higher power than the nonparametric bootstrap, a behavior that is more pronounced for the $\chi^2$-distribution (Figure~\ref{power_chi}). 
	Moreover, the power analysis shows a clear advantage of applying the parametric bootstrap approach to the MATS over its application to the WTS. For example, in the scenario with normally distributed data, $d=8$ dimensions, covariance setting S4 and $\vn = (10, 20)^\top$ observations (Figure \ref{power_n}), the parametric bootstrap MATS has twice as much power as its WTS version in case of $\delta = 0.5$ (34.4\% as compared to 16.7\%). Similar differences can also be observed in some of the other settings.

	\begin{figure}[H]
		\centering
		\includegraphics[width=0.8\textwidth]{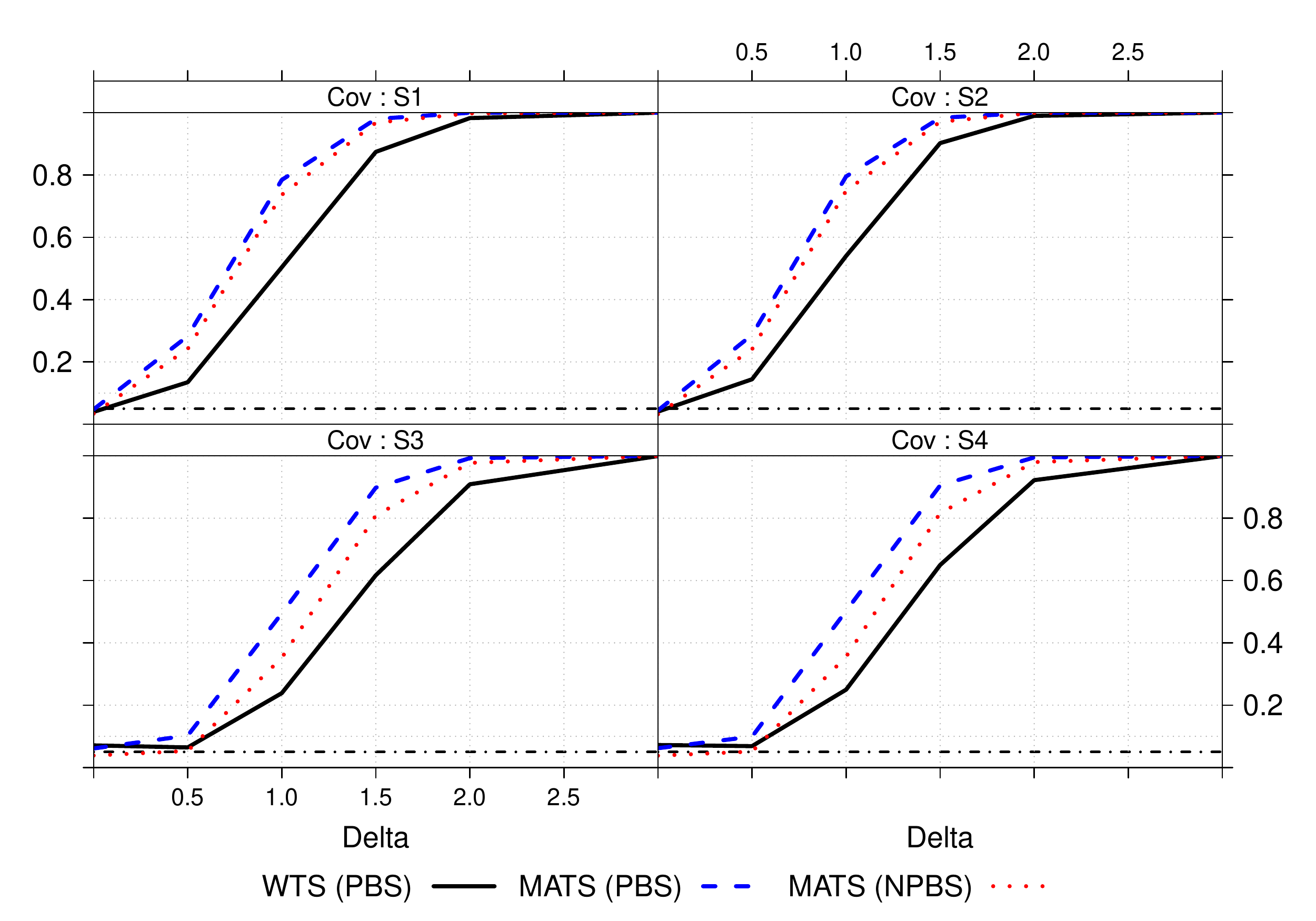}
		\caption {Empirical power results for the WTS with parametric bootstrap as well as the MATS with parametric (PBS) and nonparametric (NPBS) bootstrap for $\chi^2_3$-distributed data with $d=4$ dimensions and sample sizes $\vn=(10, 10)^\top$.}
		\label{power_chi}
	\end{figure}

	\begin{figure}[H]
		\centering
		\includegraphics[width=0.8\textwidth]{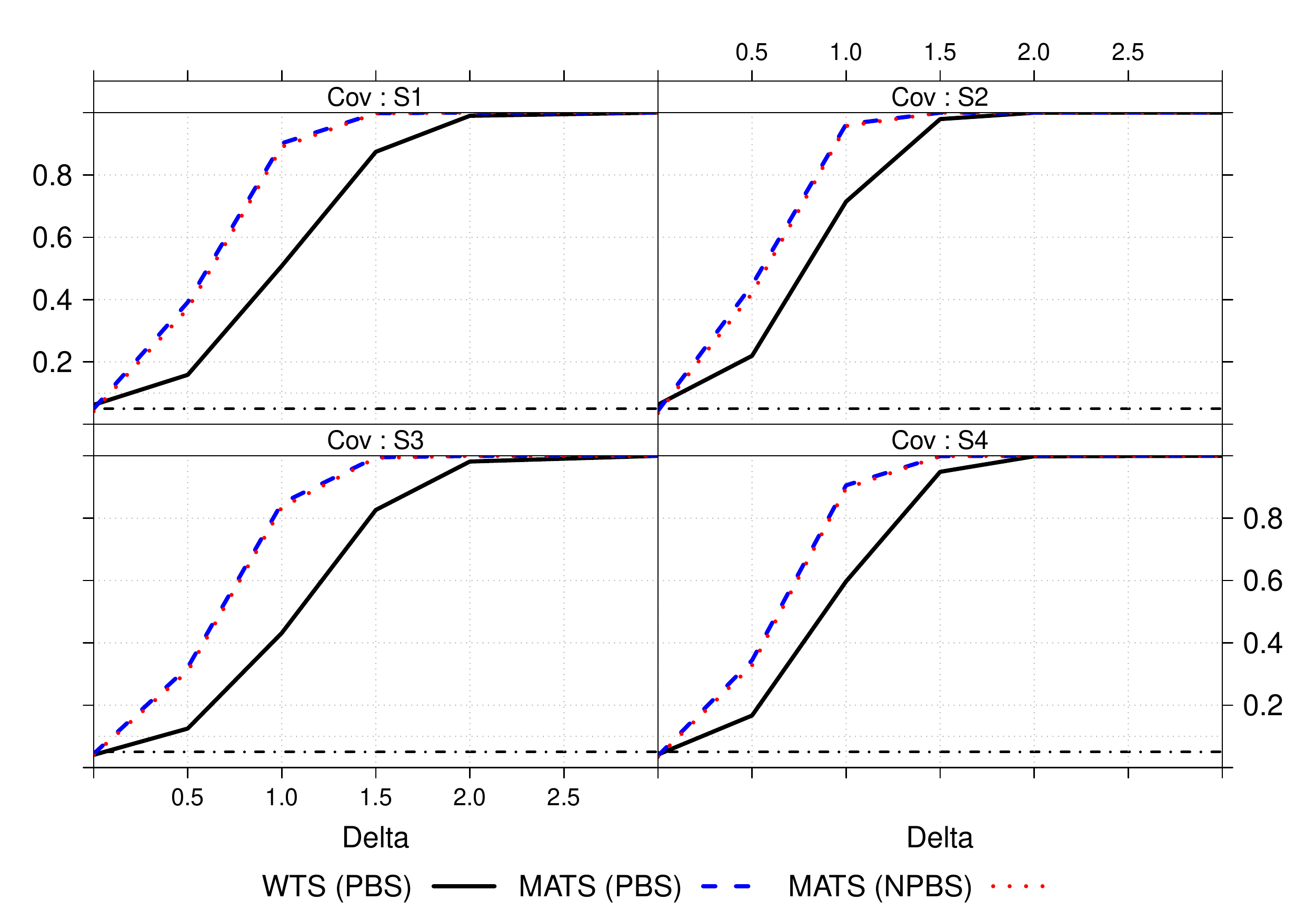}
		\caption {Empirical power results for the WTS with parametric bootstrap as well as the MATS with parametric (PBS) and nonparametric (NPBS) bootstrap for normally distributed data with $d=8$ dimensions and sample sizes $\vn=(10, 20)^\top$.}
		\label{power_n}
	\end{figure}
	
}

\section{Application: Analysis of the Data Example}\label{app}
{\color{black}
	As a data example, we consider 7 demographic factors of US citizens in 43 states. Our aim is to investigate whether these factors differ between the states.
	The full data set `county\_facts.csv` is available from kaggle (https://www.kaggle.com/joelwilson/2012-2016-presidential-elections).
	 In order to have sufficient sample sizes for the analysis and to avoid a high-dimensional setting, we exclude all states with less than 15 counties. In particular, we removed Connecticut, Delaware, Hawaii, Massachusetts, New Hampshire, Rhode Island and Vermont. We consider the following demographic factors: the population estimate for 2014 (PST045214), the percentage of female citizens in 2014 (SEX255214) as well as the percentage of white (RHI125214), black or African American (RHI225214), American Indian and Alaska native (RHI325214), Asian (RHI425214) and native Hawaiian and other pacific islanders (RHI525214) citizens in 2014. This results in a one-way layout with $a=43$ levels of the factor `state` and $d=7$ dimensions.
The sample sizes and mean values for the different states can be found in Table \ref{tab:descriptive}. Figure \ref{fig:whites} exemplarily displays boxplots for the percentage of white citizens across the different states.

\begin{sidewaystable}[ht]
	\centering
	\caption{\color{black} Descriptive statistics of the data example: Reported are the sample sizes and the 7-dimensional mean vectors for each of the 43 states.}
	\label{tab:descriptive}
	\begin{tabular}{ccccccccc}
		\hline
		State & n & PST045214 &  SEX255214 &  RHI125214 &  RHI225214 &  RHI325214 &  RHI425214 &  RHI525214 \\ 
		\hline
		 AK &  29 & 25404.55 & 45.73 & 52.51 & 1.92 & 31.89 & 5.68 & 0.55 \\ 
		 AL &  67 & 72378.76 & 51.26 & 68.26 & 28.66 & 0.80 & 0.73 & 0.11 \\ 
		 AR &  75 & 39551.59 & 50.54 & 80.41 & 16.13 & 0.89 & 0.78 & 0.10 \\ 
		 AZ &  15 & 448765.60 & 49.63 & 78.99 & 2.33 & 14.76 & 1.45 & 0.20 \\ 
		 CA &  58 & 669008.62 & 49.52 & 81.59 & 3.57 & 3.16 & 7.50 & 0.39 \\ 
		 CO &  64 & 83685.41 & 48.05 & 92.62 & 1.81 & 2.06 & 1.30 & 0.12 \\ 
		 FL &  67 & 296914.88 & 48.60 & 80.61 & 15.00 & 0.73 & 1.70 & 0.10 \\ 
		 GA & 159 & 63505.30 & 50.31 & 68.19 & 28.36 & 0.49 & 1.30 & 0.12 \\ 
		 IA &  99 & 31385.11 & 50.15 & 95.69 & 1.43 & 0.46 & 1.13 & 0.08 \\ 
		 ID &  44 & 37146.91 & 49.26 & 94.36 & 0.58 & 2.03 & 0.82 & 0.16 \\ 
		 IL & 102 & 126280.20 & 49.93 & 91.67 & 5.30 & 0.35 & 1.23 & 0.03 \\ 
		 IN &  92 & 71704.95 & 50.20 & 94.42 & 2.85 & 0.36 & 0.99 & 0.04 \\ 
		 KS & 105 & 27657.34 & 49.74 & 93.64 & 2.11 & 1.23 & 0.91 & 0.08 \\ 
		 KY & 120 & 36778.81 & 50.25 & 93.87 & 3.86 & 0.29 & 0.57 & 0.06 \\ 
		 LA &  64 & 72651.19 & 49.95 & 64.63 & 32.08 & 0.84 & 0.94 & 0.05 \\ 
		 MD &  24 & 249016.96 & 50.88 & 73.14 & 20.58 & 0.44 & 3.41 & 0.10 \\ 
		 ME &  16 & 83130.56 & 50.89 & 95.67 & 0.98 & 0.87 & 0.89 & 0.02 \\ 
		 MI &  83 & 119396.11 & 49.67 & 91.18 & 4.05 & 1.69 & 1.03 & 0.03 \\ 
		 MN &  87 & 62726.13 & 49.88 & 92.93 & 1.59 & 2.24 & 1.47 & 0.06 \\ 
		 MO & 115 & 52726.86 & 50.11 & 93.08 & 3.71 & 0.62 & 0.74 & 0.12 \\ 
		 MS &  82 & 36513.16 & 51.01 & 56.48 & 41.21 & 0.68 & 0.57 & 0.04 \\ 
		MT &  56 & 18278.20 & 49.16 & 88.81 & 0.41 & 8.00 & 0.48 & 0.05 \\ 
		NC & 100 & 99439.64 & 50.82 & 74.15 & 20.80 & 1.93 & 1.26 & 0.10 \\ 
		ND &  53 & 13952.49 & 48.69 & 90.26 & 0.79 & 6.79 & 0.59 & 0.04 \\ 
		 NE &  93 & 20231.22 & 49.82 & 95.43 & 0.93 & 1.73 & 0.59 & 0.07 \\ 
		 NJ &  21 & 425627.38 & 51.06 & 76.97 & 13.19 & 0.56 & 7.16 & 0.10 \\ 
		 NM &  33 & 63199.15 & 49.37 & 86.02 & 1.83 & 8.70 & 1.11 & 0.14 \\ 
		 NV &  17 & 167005.82 & 47.41 & 87.34 & 2.88 & 4.27 & 2.26 & 0.31 \\ 
		 NY &  62 & 318487.53 & 50.22 & 87.43 & 6.98 & 0.70 & 2.90 & 0.06 \\ 
		 OH &  88 & 131751.85 & 50.38 & 92.70 & 4.33 & 0.28 & 0.97 & 0.02 \\ 
		 OK &  77 & 50364.30 & 49.84 & 78.41 & 3.78 & 11.18 & 0.86 & 0.14 \\ 
		 OR &  36 & 110284.42 & 50.04 & 91.49 & 0.89 & 2.48 & 1.77 & 0.30 \\ 
		 PA &  67 & 190853.87 & 50.03 & 91.99 & 4.86 & 0.27 & 1.46 & 0.03 \\ 
		 SC &  46 & 105053.96 & 50.93 & 60.75 & 36.10 & 0.64 & 0.93 & 0.09 \\ 
		 SD &  66 & 12926.89 & 49.48 & 82.51 & 0.78 & 14.02 & 0.63 & 0.04 \\ 
		 TN &  95 & 68940.55 & 50.46 & 89.75 & 7.50 & 0.45 & 0.73 & 0.05 \\ 
		 TX & 254 & 106129.76 & 49.20 & 89.24 & 6.82 & 1.17 & 1.16 & 0.08 \\ 
		UT &  29 & 101479.38 & 49.12 & 92.88 & 0.66 & 3.30 & 1.04 & 0.36 \\ 
		 VA & 134 & 62136.49 & 50.25 & 75.67 & 18.84 & 0.51 & 2.09 & 0.07 \\ 
		 WA &  39 & 181064.87 & 49.85 & 88.87 & 1.61 & 3.01 & 2.79 & 0.34 \\ 
		 WI &  72 & 79966.17 & 49.64 & 92.55 & 1.71 & 2.96 & 1.31 & 0.04 \\ 
		 WV &  55 & 33642.29 & 50.12 & 95.57 & 2.38 & 0.25 & 0.49 & 0.01 \\ 
		 WY &  23 & 25397.96 & 49.04 & 94.04 & 1.17 & 2.13 & 0.83 & 0.10 \\ 
		\hline
	\end{tabular}
\end{sidewaystable}

\begin{figure}[h]
		\centering
			\includegraphics[width = \textwidth]{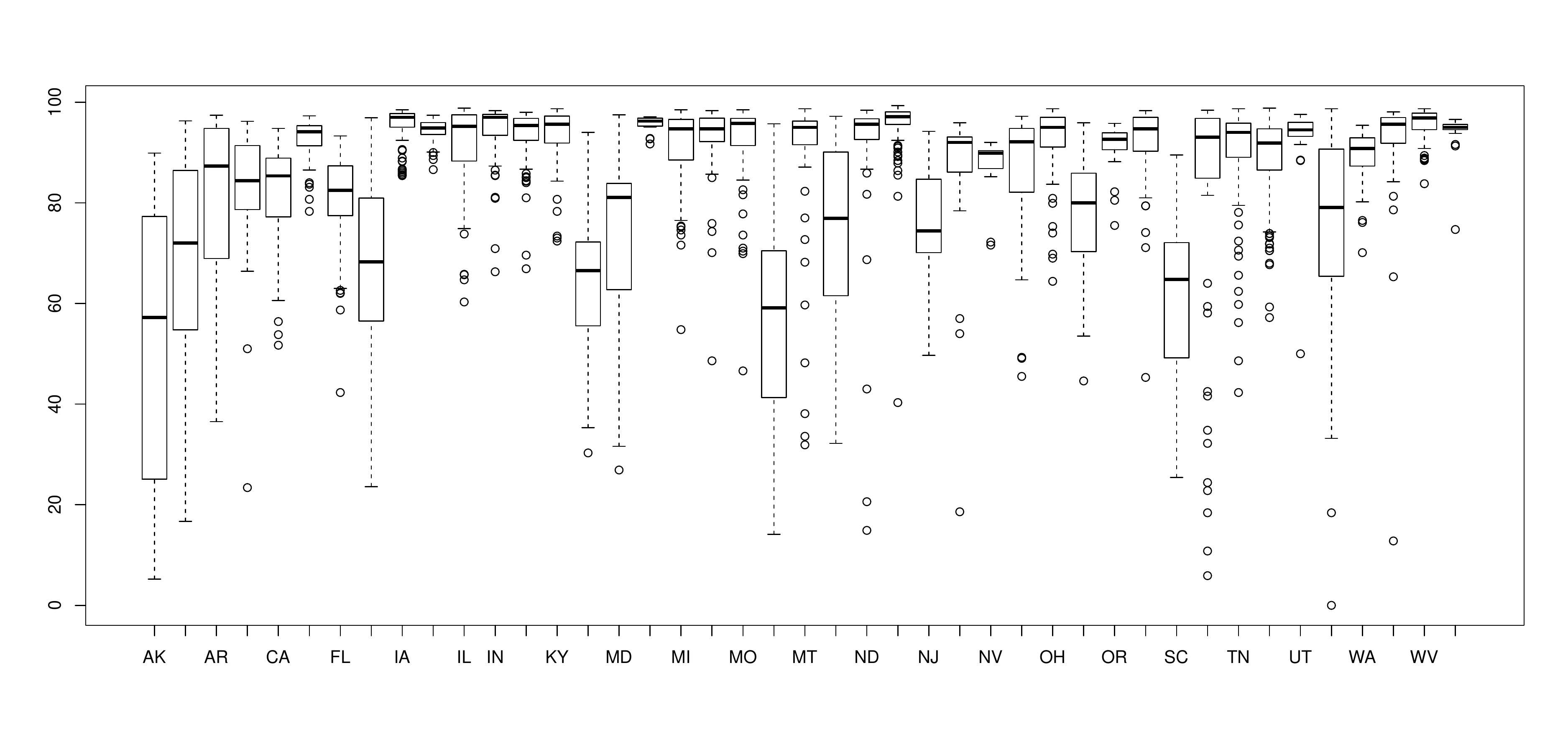}
			\caption{Boxplots of the percentage of white citizens across the different states.}
			\label{fig:whites}
\end{figure}

We now want to analyze, whether there is a significant difference in the multivariate means for the different states. The null hypothesis of interest thus is $H_0: \{(\vP_{43} \kp \vI_7) \vmu = \vnull\}$.
Since the empirical covariance matrix is computationally singular in this example (reciprocal condition number 1.7e-16), we cannot apply the Wald-type test. Thus, we consider the parametric bootstrap approach of the MATS which yielded the best results in the simulation study.
Computation of the MATS results in a value of $Q_N = 393.927$ and the parametric bootstrap routine with 1,000 bootstrap runs gives a $p$-value of $p < 0.0001$, implying that there is indeed a significant difference between the states with respect to the 7 demographic measurements.

A confidence region for this effect can be constructed as described in Section \ref{sec:stats}. The analysis of this example, including the calculation of the confidence region, can be conducted using the \textsf{R} package \texttt{MANOVA.RM}.

}

\section{Conclusions and Discussion} \label{dis}

We have investigated a test statistic for multivariate data (MATS) which is based on a modified Dempster statistic. 
Contrary to classical MANOVA models, we incorporate general heteroscedastic designs and allow for singular covariance matrices 
while postulating their existence as solely distributional assumption. Moreover, our proposed MATS statistic is invariant under linear transformations of the response variables. 

In order to improve the small sample behavior of the test statistic, we have investigated different bootstrap approaches, namely a parametric bootstrap, a wild bootstrap {\color{black} and a nonparametric bootstrap} procedure. We have rigorously proven that {\color{black}they} lead to asymptotically exact and consistent tests and even analyzed their local power behavior.\\
In a large simulation study, the parametric bootstrap turned out to perform {\color{black} best} in most scenarios, even with skewed data and heteroscedastic variances. Although the type-I error control is still not ideal in the latter case, the method performed advantageous over the parametric bootstrap of the WTS proposed in \cite{Kon:2015} and has the additional advantage of being applicable to situations with singular covariance matrices. In situations with skewed distributions, the parametric bootstrap of the MATS yielded more robust results than the WTS. The wild bootstrap approach, in contrast, turned out to be very {\color{black} liberal in all scenarios, while the nonparametric bootstrap was mostly slightly more conservative than the parametric bootstrap. Power simulations showed a clear advantage of the parametric bootstrap MATS compared to the WTS (PBS) as well as the nonparametric bootstrap.} {\color{black} All in all, we therefore} recommend the parametric bootstrap based on the MATS {\color{black} for practical applications in a multivariate setting.}\\
Furthermore, we have constructed confidence regions and simultaneous confidence intervals for contrasts $\vh^\top\vmu$ based on the bootstrap quantiles. These confidence regions provide an additional benefit for the analysis of multivariate data since they allow for more detailed insight into the nature of the estimates.\\
In order to facilitate application of the proposed methods, the parametric bootstrap test and the calculation of confidence regions are implemented in the \textsf{R} package \texttt{MANOVA.RM}. 

Following the idea of \cite{srivastava2013tests} we plan to extend our concepts to the high-dimensional setting, i.e., where the sample size $N$ may be less than the dimension $d$. This approach looks promising, since we have seen in the simulation study that the MATS with the parametric bootstrap approach exhibited an improved type-I error control with increasing $d$. However, the extension to high-dimensional data requires different techniques and will be part of future research.

\section*{Acknowledgment}
The authors would like to thank Dr.~Jan Paul and Prof.~Dr.~Volker Rasche for providing the {\color{black}cardiology} data example {\color{black}used in the supplement}.
This work was supported by the German Research Foundation projects DFG PA 2409/3-1 and PA 2409/4-1.




\section*{References}
\bibliography{Literatur}{}
\bibliographystyle{abbrv}

\newpage

\title{Supplementary Material to\\
	MATS: Inference for potentially Singular and Heteroscedastic MANOVA
}

\author{Sarah Friedrich\corref{cor1}}
\ead{sarah.friedrich@uni-ulm.de}
\author{Markus Pauly\corref{cor1}}

\address{ Institute of Statistics, Ulm University, Germany}

\begin{abstract}
	In this supplementary material to the authors' paper "MATS: Inference for potentially singular and heteroscedastic MANOVA" we provide the proofs of all theorems as well as some additional simulation results {\color{black} for different distributions and a two-way layout}. 
	Furthermore, we {\color{black} provide an additional data example from cardiology, where we also explain the problem of the ATS in more detail.}
	
\end{abstract}

\begin{keyword}
	Multivariate Data; Parametric Bootstrap; Confidence Regions; Singular Covariance Matrices
\end{keyword}

\maketitle

\setcounter{section}{7}
\setcounter{table}{6}

\newpage


\section{Proofs}

{\bf Proof of Theorem \ref{theo:MATS}}\\
The result follows directly from the representation theorem for quadratic forms \citep{Rao} and the continuous mapping theorem by noting that $\sqrt{N}(\volX_{\bullet}-\vmu)$ has, asymptotically, as $N \to \infty$, a multivariate normal distribution with mean $\vnull$ and covariance matrix $\vSigma = \lim_{N \to \infty} \vSigma_N= \diag(\kappa_i^{-1} \vV_i)$. Moreover, $\vDh_N$ is consistent for $\vD = \diag(\kappa_i^{-1} \sigma_{is}^2)$, where the latter is of full rank by assumption. 
Thus, $\vT \vDh_N\vT$ converges in probability to $\vT\vD\vT$ and since there is finally no rank jump in this convergence we eventually obtain
\bqan \label{consMPI}
(\vT \vDh_N\vT)^+ \xrightarrow{\Pr}(\vT\vD\vT)^+,
\eqan
where $\xrightarrow{\Pr}$ denotes convergence in probability.\qed

{\bf Proof of Theorem \ref{theo:MATS_pbs}}\\
Let $$ \vY_{ik}:= \vX_{ik}\cdot \1\{||\vX_{ik}||\leq \delta\}$$ for $\delta >0$. Then, $\vY_{ik}$ has finite moments of any order, especially fourth moments exist.
Analogously, given $\vY=(\vY_{ik})_{i,k}$, let $\vY_{ik}^*  \stackrel{\textrm{i.i.d.}}{\sim} \mathcal{N}(0, \vwhV_i(\delta))$, where $\vwhV_i(\delta) = (n_i-1)^{-1}\sum_{k=1}^{n_i}(\vY_{ik}-\volY_{i\cdot})(\vY_{ik}-\volY_{i\cdot})^\top$ and define $Q_{N, \delta}^* = N (\volY_{\bullet}^*)^\top \vT (\vT \vwhD_{\delta}^* \vT)^+ \vT \volY_{\bullet}^*$, where $\vwhD^*_\delta = \diag(N/n_i \cdot (\hsigma_{is}^*(\delta))^2)$ and $(\hsigma_{is}^*(\delta))^2$ is the empirical variance of $Y_{iks}^*$.

First, we apply the multivariate Lindeberg-Feller CLT to show that, given $\vX$, $\sqrt{N} ~ \volY^*_{\bullet}$ converges in distribution to a normal distributed random variable.
Therefore, consider $\widetilde{\vY_i}^* := \sqrt{N}/n_i \vY_{ik}^*$ and $\vV_i(\delta) = \Cov(\vY_{ik})$. 
The Lindeberg-Feller CLT now yields convergence in distribution given the data $\vX$ to a normal distributed random variable

\bqan \label{LFCLT}
\sqrt{N}~\volY_{\bullet}^* \xrightarrow{\mathcal{D}|\vX} \mathcal{N}(\vnull, \vSigma_\delta),
\eqan
if we proof the following conditions:
\begin{eqnarray}
	\sum_{i=1}^a \sum_{k=1}^{n_i} \E(\widetilde{\vY_i}^*|\vX)&=&0 \label{LF1}\\
	\bigoplus_{i=1}^a \sum_{k=1}^{n_i} \Cov(\widetilde{\vY_i}^*|\vX) &\xrightarrow{\Pr}& \vSigma_\delta := \bigoplus_{i=1}^a \frac{1}{\kappa_i} \vV_i(\delta) \label{sigdelta}\\
	\sum_{i=1}^a \sum_{k=1}^{n_i} \E \left(||\widetilde{\vY_i}^*||^2 \cdot  \1\{||\widetilde{\vY_i}^*||>\epsilon\}|\vX \right) &\xrightarrow{\Pr}& 0 ~~\forall ~ \epsilon>0. \label{LF3}
\end{eqnarray}
Condition \eqref{LF1} follows since, given the data, $\vY_{ik}^*  \stackrel{\textrm{i.i.d.}}{\sim} \mathcal{N}(0, \vwhV_i(\delta))$. For condition \eqref{sigdelta}, note that
\begin{eqnarray*}
	\bigoplus_{i=1}^a \sum_{k=1}^{n_i} \Cov(\widetilde{\vY_i}^*|\vX) = \bigoplus_{i=1}^a \sum_{k=1}^{n_i} \Cov \left(\frac{\sqrt{N}}{n_i}{\vY_{ik}}^*|\vX \right) =\bigoplus_{i=1}^a \sum_{k=1}^{n_i} \frac{N}{n_i^2}\vwhV_i(\delta)  =\bigoplus_{i=1}^a \frac{N}{n_i}\vwhV_i(\delta).
\end{eqnarray*}
Thus, since $\vwhV_i(\delta)$ is a consistent estimator of $\vV_i(\delta)$, \eqref{sigdelta} follows. For \eqref{LF3} note that 
$$
\1\{||\widetilde{\vY_i}^*||>\epsilon\} = 1 \Leftrightarrow  \delta \geq ||\vY_{ik}^*|| > \frac{n_i}{\sqrt{N}}\epsilon = \frac{n_i}{N}\sqrt{N}\epsilon.
$$
Since $ n_i/N \to \kappa_i>0$, the right hand side converges to infinity as $N \to \infty$. Therefore, for arbitrary fixed $\delta>0$ and $\epsilon>0$ we finally have 
$\1\{||\widetilde{\vY_i}^*||>\epsilon\} = 0$ for $N$ large enough and Equation \eqref{LF3} follows. Altogether this proves \eqref{LFCLT}.

Since $\vwhD_\delta^* \xrightarrow{\Pr} \vD_\delta = \diag(\kappa_i^{-1} \Var(Y_{iks})), i=1, \dots, a, ~ s = 1, \dots, d$ due to existence of finite fourth moments of the truncated random variables $\vY$, it now follows from continuous mapping and the representation theorem for quadratic forms that
$$
Q_{N, \delta}^* =  N (\volY_{\bullet}^*)^\top \vT (\vT \vwhD_{\delta}^* \vT)^+ \vT \volY_{\bullet}^* \xrightarrow{\mathcal{D}|\vX} \tilde{Z}, ~ N \to \infty,
$$
in probability, see, e.g., \cite{Kon:2015} and the references cited therein for similar arguments. Here, $\tilde{Z}=\sum_{i=1}^a \sum_{s=1}^d \tilde{\lambda}_{is} \tilde{Z}_{is}$ with $\tilde{Z}_{is} \sim \chi^2_1$ and $\tilde{\lambda}_{is}$ are the eigenvalues of $\vT (\vT \vD_{\delta} \vT)^+ \vT \vSigma_{\delta}$. 
Furthermore, since
$$
\Cov(\vY_{ik})=\Cov(\vX_{ik}\cdot \1\{||\vX_{ik}|| \leq \delta\}) \to \Cov(\vX_{ik}), ~\delta \to \infty ,
$$
by dominated convergence and analogously $\vD_\delta \to \vD, \delta \to \infty$, it follows that
$$
\tilde{Z} \xrightarrow{\mathcal{D}|\vX} Z, ~\delta \to \infty
$$
in probability, where $Z$ is the limit variable of $Q_N$ given in Theorem \ref{theo:MATS}.
Thus, it remains to show that 
$$\lim_{\delta \to \infty} \limsup_{N \to \infty} \Pr(|Q_{N, \delta}^* - Q_N^*|>\epsilon| \vX) \stackrel{\textrm{a.s.}}{=} 0 ~ \textrm{for all}~ \epsilon>0.$$

Let $\vwtY:= \sqrt{N} \vT \volY_{\bullet}^*$, $\vwtX:= \sqrt{N} \vT \volX_{\bullet}^*$, $\vM_{\delta} := (\vT\vD_{\delta}^* \vT)^+$ and $\vM := (\vT \vD^* \vT)^+$. Then,  $Q_{N, \delta}^* = \widetilde{\vY}^\top\vM_{\delta}\widetilde{\vY}$ and $Q_N^* = \widetilde{\vX}^\top\vM \widetilde{\vX}$ and therefore
$$
Q_{N, \delta}^* - Q_N^* = \underbrace{\widetilde{\vY}^\top\vM_{\delta}(\widetilde{\vY}-\widetilde{\vX})}_{(A)}+\underbrace{(\widetilde{\vY}^\top\vM_{\delta}- \widetilde{\vX}^\top\vM)\widetilde{\vX}}_{(B)}.
$$

First, consider part $(A)$ and let $\boldsymbol{\xi}_{ik} := \vX_{ik} - \vY_{ik} = \vX_{ik} \1\{||\vX_{ik}||>\delta\}$. Another application of the multivariate Lindeberg-Feller CLT shows that 
$$
\sqrt{N}(\volY_{\bullet}^* - \volX_{\bullet}^*) \xrightarrow{\mathcal{D}|\vX} \mathcal{N}(\vnull, \widetilde{\vSigma}_\delta)
$$
in probability, where $\widetilde{\vSigma}_\delta :=\bigoplus_{i=1}^a \kappa_i^{-1} \Cov(\vX_{ik} \1\{||\vX_{ik}||> \delta\})=\bigoplus_{i=1}^a \kappa_i^{-1}\Cov(\boldsymbol{\xi}_{ik})$.\\
Thus, 
$$
\widetilde{\vY} - \widetilde{\vX} \xrightarrow{\mathcal{D}|\vX}\mathcal{N}(0, \vT \widetilde{\vSigma}_\delta)
$$
in probability and the representation theorem again yields $\widetilde{\vY}^\top\vM_{\delta}(\widetilde{\vY}-\widetilde{\vX}) \xrightarrow{\mathcal{D}|\vX} B_\delta = \sum_{i=1}^a \sum_{s=1}^d \eta_{is}^{(\delta)} B_{is}^2$ in probability, where $B_{is}^2 \sim \chi^2_1$ and $\eta_{is}^{(\delta)}$ are the eigenvalues of $(\vT \vSigma_\delta)^{1/2} \vM_\delta (\vT \vwtSigma_\delta)^{1/2}$.\\
By dominated convergence it follows that $\vwtSigma_\delta \to 0$ for $\delta \to \infty$. Since $\vSigma_\delta \to \vSigma$ and $\vD_\delta \to \vD$ we finally obtain 
$B_\delta \to 0$ as $\delta \to \infty$. Altogether, this proves
$$
\lim_{\delta \to \infty} \limsup_{N \to \infty} (A) = 0.
$$
It remains to consider part $(B)$ which we expand as
$$
(\widetilde{\vY}^\top\vM_{\delta}- \widetilde{\vX}^\top\vM)\widetilde{\vX} = [\widetilde{\vY}^\top\vM_{\delta}- \widetilde{\vX}^\top\vM_\delta + \widetilde{\vX}^\top (\vM_\delta - \vM)]\widetilde{\vX}.
$$
Using similar arguments as above, it follows that given the data
$$
(\widetilde{\vY}^\top- \widetilde{\vX}^\top)\vM_\delta \widetilde{\vX} 
$$
converges to 0 in probability for $N \to \infty$ and, subsequently, $\delta \to \infty$. 

Finally, $(\hat{\sigma}^*_{is}(\delta))^2 - (\hat{\sigma}^*_{is})^2$ converges to zero (where $(\hat{\sigma}^*_{is})^2$ is the empirical variance of $X_{iks}^*$) by 
dominated convergence and consistency of the variance estimators and it follows that $\vM_\delta -\vM$ converges to 0. This concludes the proof.
\qed
\\
{\bf Proof of Theorem \ref{theo:MATS_wild}}\\
Analogous to the proof of Theorem \ref{theo:MATS_pbs}, we define
$$ \vY_{ik}:= \vX_{ik}\cdot \1\{||\vX_{ik}||\leq \delta\}$$ for $\delta >0$ as well as, given $\vY$, $\vY_{ik}^\star = W_{ik}(\vY_{ik}-\volY_{i\cdot})$ and $Q_{N, \delta}^\star = N (\volY_{\cdot}^\star)^\top \vT (\vT \vwhD_{\delta}^\star \vT)^+ \vT \volY_{\cdot}^\star$.

The first part of the proof follows analogous to the proof of Theorem \ref{theo:MATS_pbs} above.
It remains to show that 
$
(\hat{\sigma}^\star_{is}(\delta))^2 - (\hat{\sigma}^\star_{is})^2
$
converges to zero. Therefore, consider 
$$
(*) := (\hat{\sigma}^\star_{is}(\delta))^2 - (\hat{\sigma}^\star_{is})^2 = \frac{1}{n_i}\sum_{k=1}^{n_i} W_{ik}^2 \xi_{iks}^2 - (\olY_{i\cdot s}^\star)^2+(\olX_{i\cdot s}^\star)^2,
$$
where again $\xi_{iks} := X_{iks} - Y_{iks}$.
For the first summand on the right hand side it holds
$$
\E \left(\frac{1}{n_i}\sum_{k=1}^{n_i} W_{ik}^2 \xi_{iks}^2 | \vX \right) = \frac{1}{n_i}\sum_{k=1}^{n_i} \xi_{iks}^2 \xrightarrow[N \to \infty]{\textrm{a.s.}} \E(X_{i1s}^2 \1\{||\vX_{ik}||>\delta\}).
$$
Now letting $\delta \to \infty$, it follows from dominated convergence that
$$
\E(X_{i1s}^2 \1\{||\vX_{ik}||>\delta\}) \xrightarrow{\delta \to \infty} 0.
$$

Concerning $(\olX_{i\cdot s}^\star)^2-(\olY_{i\cdot s}^\star)^2$, we first consider $\olX_{i\cdot s}^\star-\olY_{i\cdot s}^\star=n_i^{-1}\sum_{k=1}^{n_i} W_{ik} \xi_{iks}$.
It holds that $\E(\olX_{i\cdot s}^\star-\olY_{i\cdot s}^\star| \vX)=0$ as well as $\Var(\olX_{i\cdot s}^\star-\olY_{i\cdot s}^\star|\vX)\xrightarrow{N \to \infty} \Var(\olX_{i \cdot s} - \olY_{i \cdot s})  \xrightarrow{\delta \to \infty}0$ and therefore
$$
\lim_{\delta \to \infty} \limsup_{N \to \infty}  \Pr(|\olX_{i\cdot s}^\star-\olY_{i\cdot s}^\star|>\epsilon|\vX) = 0.
$$
The continuous mapping theorem now implies
$$
\lim_{\delta \to \infty} \limsup_{N \to \infty}  \Pr((\olX_{i\cdot s}^\star-\olY_{i\cdot s}^\star)^2>\epsilon|\vX) = 0
$$
and furthermore
$$
(\olX_{i\cdot s}^\star-\olY_{i\cdot s}^\star)^2 \1\{\olX_{i\cdot s}^\star > \olY_{i\cdot s}^\star\}\xrightarrow{\Pr}0.
$$
Therefore,
$$
0 \leq \left((\olX_{i\cdot s}^\star)^2-(\olY_{i\cdot s}^\star)^2\right)\1\{\olX_{i\cdot s}^\star> \olY_{i\cdot s}^\star\} \to 0
$$
and analogously
$$
0 \leq \left((\olX_{i\cdot s}^\star)^2-(\olY_{i\cdot s}^\star)^2\right)\1\{\olX_{i\cdot s}^\star< \olY_{i\cdot s}^\star\} \to 0
$$
and therefore
$$
\lim_{\delta \to \infty} \limsup_{N \to \infty}  \Pr(|(\olX_{i\cdot s}^\star)^2-(\olY_{i\cdot s}^\star)^2|>\epsilon|\vX) = 0.
$$

Altogether, this implies by the general Markov inequality that
$$
\lim_{\delta \to \infty} \limsup_{N \to \infty} \Pr((*)> \epsilon|\vX) \leq \lim_{\delta \to \infty} \limsup_{N \to \infty} \frac{1}{\epsilon}\E((*)|\vX) \stackrel{\textrm{a.s.}}{=} 0,
$$
which concludes the proof.
\qed

{\color{black}
	{\bf Proof of Theorem \ref{theo:MATS_NPBS}}\\
	The result for the nonparametric bootstrap can be proved by conditionally following the lines of the proof of Theorem \ref{theo:MATS}. First note, that conditional independence of the bootstrap sample 
	$\vX_{ik}^\dagger, i=1,\dots,a,\, k=1,\dots,n_i$ together with the multivariate CLT for the bootstrap given in \cite{bickel1981some} implies that the conditional distribution of $\sqrt{N}(\volX_{\bullet}^\dagger-\volX_{\bullet})$ asymptotically, as $N \to \infty$, coincides with a multivariate normal distribution with mean $\vnull$ and covariance matrix $\vSigma = \diag(\kappa_i^{-1} \vV_i)$ (almost surely). Moreover, the law of large numbers for the bootstrap (see, e.g., 
	\cite{csorgo1992law}) implies that $\vDh_N^\dagger$ converges almost surely to $\vD = \diag(\kappa_i^{-1} \sigma_{is}^2)$. 
	We can thus conclude as in the proof of Theorem \ref{theo:MATS}: $\vT \vDh_N^\dagger\vT \to \vT\vD\vT$ holds almost surely and since there is finally no rank jump in this convergence we also obtain $(\vT \vDh_N^\dagger\vT)^+ \xrightarrow{\Pr}(\vT\vD\vT)^+$ almost surely. Putting these ingredients together with the continuous mapping theorem finally proves the convergence.
	\qed
}

{\bf Proofs of the results in Section \ref{tests}}\\
Theorems \ref{theo:MATS_pbs} -- \ref{theo:MATS_NPBS} directly imply that the corresponding bootstrap tests $\varphi^*=\1\{Q_N > c_{1-\alpha}^*\}$ asymptotically keep the pre-assigned level $\alpha$, since $c_{1-\alpha}^*$ is the $(1-\alpha)$-quantile of the (conditional) bootstrap distribution, which, given the data, converges weakly to the null distribution of $Q_N$ in probability.\\
For local alternatives $H_1: \vT \vmu = \sqrt{N}^{-1}\vT \vnu, \vnu \in \mathbb{R}^{ad}$,
it holds that
$$
\sqrt{N}\vT \volX_{\bullet} = \sqrt{N}\vT(\volX_{\bullet}-\vmu) + \vT \vnu
$$
has, asymptotically, as $N \to \infty$, a multivariate normal distribution with mean $\vT \vnu$ and covariance matrix $\vT \vSigma \vT$ and thus $Q_N$ converges to $\zeta (\vT \vD \vT)^+ \zeta$, where $\zeta \sim \mathcal{N}(\vT \vnu, \vT \vSigma \vT)$, by using \eqref{consMPI} again.
Thus, Theorems \ref{theo:MATS_pbs} -- \ref{theo:MATS_NPBS} imply that the bootstrap tests have the same asymptotic power as $\varphi_N = \1\{Q_N> c_{1-\alpha}\}$ and the asymptotic relative efficiency of the bootstrap tests $\varphi^*$ compared to $\varphi_N$ is 1 in this situation.
\qed

{\color{black}
	{\bf Proof of the results in Section \ref{confint}}\\
	In order to derive a simultaneous contrast test formulated in the maximum statistic, we need to analyze the asymptotic joint distribution of the vector of test statistics $\vQ =(Q_N^1, \dots, Q_N^q)^\top$. First note that $\vQ$ can be re-written as
	\bqan \label{eq:Q}
	\sqrt{N} (\vH \volX_\bullet - \vH \vmu)^\top \sqrt{N} \diag \left(\vh_\ell^\top \volX_\bullet \right) \diag \left((\vh_\ell^\top \vwhD_N \vh_\ell)^{-1}\right).
	\eqan
	The last diagonal matrix converges to $\diag\left((\vh_\ell^\top \vD \vh_\ell)^{-1}\right)$, while the first part can be viewed as $\psi \left(\sqrt{N} (\vH \volX_\bullet - \vH \vmu)\right)$ for a continuous function $\psi$. Due to the results above, $\sqrt{N} (\vH \volX_\bullet - \vH \vmu)$ converges to a multivariate normally distributed random vector $\Xi$ with mean $\vnull$ and covariance matrix $\vH \vSigma \vH^\top$. Thus, \eqref{eq:Q} converges in distribution to $\psi(\Xi) \cdot \diag\left((\vh_\ell^\top \vD \vh_\ell)^{-1}\right)$ due to the continuous mapping theorem and Slutzky. The same distributional convergence also holds for the bootstrapped test statistic $\vQ^*=(Q_N^{1, *}, \dots, Q_N^{q,*})^\top$ given the data in probability due to Theorems \ref{theo:MATS_pbs} -- \ref{theo:MATS_NPBS} (they imply that
	$\widehat{\sigma}_{is}^*$ are consistent estimates for ${\sigma}_{is}$ ($i=1,\dots,a, \, s=1,\dots,d$) and that 
	$ \sqrt{N} \vH\volX_\bullet^*$ converges to $\Xi$ in distribution given the data in probability). Since $\max$ is continuous, it thus follows that
	$$
	\sup_x |\Pr\{\max(Q_N^1, \dots, Q_N^q) \leq x\} - \Pr\{\max(Q_N^{1,*}, \dots, Q_N^{q, *})\leq x | \vX\} | \stackrel{\Pr}{\to} 0,
	$$
	which concludes the proof.
	\qed
	
	In future research it will be investigated which method performs preferably for the derivation of simultaneous confidence intervals.
	
}

\section{Further simulation results}

\subsection{One-way layout}

{\color{black} In this section, we present some additional simulation results for different distributions. The simulation scenarios are the same as in the paper, but we have excluded the WTS with $\chi^2$-approximation, the wild and the nonparametric bootstrap of the MATS here.}

The results are displayed in Tables \ref{app:logn} -- \ref{app:dexp} for the lognormal, $t_3$ and double-exponential distribution, respectively. The parametric bootstrap of the MATS keeps the pre-assigned $\alpha$-level very well for the $t_3$ and the double-exponential distribution. 
Note that the validity of the parametric bootstrap of the WTS has not yet been proven for the $t_3$ distribution, since fourth moments do not exist in this case.
With lognormally distributed data and negative pairing (setting 3 and 4 with $\vn =(20, 10)^\top$), all procedures show a liberal behavior. 

\subsection{Two-way layout}
To analyze the behavior of our methods in a setting with two crossed factors $A$ and $B$, we considered the following simulation design, which is again adapted from \cite{Kon:2015}. We simulated a $2\times2$ design with sample sizes $\vn^{(1)}=(n_{11}^{(1)}, n_{12}^{(1)}, n_{21}^{(1)}, n_{22}^{(1)})^\top=(7, 10, 13, 16)^\top, \vn^{(2)}=(10, 10, 10, 10)^\top,\vn^{(3)}=(16, 13, 10, 7)^\top,\vn^{(4)}=(20, 20, 20, 20)^\top$. The covariance settings were chosen similar to the one-way layout above as:

\bqa
\text{Setting 8: }&& \vV_{ij} = \vI_d + 0.5 (\vJ_d-\vI_d), ~ i, j=1, 2,\\
\text{Setting 9: }&& \vV_{ij} = \left((0.6)^{|r-s|}\right)_{r, s =1}^d, ~ i, j=1, 2,\\
\text{Setting 10: }&& \vV_{ij} = \vI_d\cdot \ell + 0.5 (\vJ_d-\vI_d), ~ \ell=1, \dots, 4,\\
\text{Setting 11: }&& \vV_{ij} = \left((0.6)^{|r-s|}\right)_{r, s =1}^d+ \vI_d \cdot \ell, ~ \ell=1, \dots, 4.\\
\eqa
Again, setting 10 and 11 combined with sample sizes $\vn^{(2)}$ and $\vn^{(3)}$ represent settings with positive and negative pairing, respectively. In this scenario, we consider three different null hypotheses of interest:
\bit
\item[(1)] The hypothesis of \emph{no effect of factor $A$}
\bqa
H_0^{\mu}(A) &:&  \{\volmu_{1 \cdot }= \volmu_{2 \cdot}\} = \{ \vH_A\vmu = \boldsymbol{0}\},
\eqa
\item[(2)] The hypothesis of \emph{no effect of factor $B$}
\bqa
H_0^{\mu}(B) &:& \{\volmu_{\cdot 1}= \volmu_{ \cdot 2}\} = \{\vH_B\vmu = \boldsymbol{0}\},
\eqa
\item[(3)] The hypothesis of \emph{no $A \times B$ interaction effect}
\bqa
H_0^{\mu}(AB) &:&  \{(\vP_a \kp \vP_b \kp \vI_d)\vmu
= \boldsymbol{0}\},
\eqa
where $ \vH_A = \vP_a \kp b^{-1} \vJ_b \kp \vI_d$ and $\vH_B = a^{-1} \vJ_a \kp \vP_b \kp \vI_d$.
\eit
The simulation results for factor $A$ and $B$ {\color{black} as well as the interaction between the factors} are in Tables \ref{app:facA} -- \ref{app:interaction}, respectively. 
{\color{black} Due to the larger total sample size $N$ in this scenario, the asymptotic results come into play and therefore all methods lead to more accurate results than in the one-way layout.
	The behavior of the tests is similar to the one-way layout: Both MATS and WTS with the parametric bootstrap approach control the type-I error accurately in most scenarios. The $\chi^2$-approximation of the WTS is very liberal again, while the wild bootstrap MATS also shows a slightly liberal behavior. 
	In situations with negative pairing (covariance setting 10 and 11 with sample size vector $\vn^{(3)}$), the parametric bootstrap MATS improves the slightly liberal behavior of the WTS, see e.g., Table \ref{app:facA} for the normal distribution, where the WTS (PBS) leads to a type-I error of 6.1\%, while the MATS (PBS) is at 4.9\%. The nonparametric bootstrap is again slightly more conservative than the parametric bootstrap, see, e.g., Table \ref{app:facB} with $\chi^2_3$-distribution and covariance settings S10 and S11. For the interaction hypothesis with $\chi^2_3$-distribution, the WTS (PBS), MATS (PBS) and MATS (NPBS) show more conservative results than for the hypotheses about the main effects. 
}

\begin{table}[H]
	\centering
	\caption{Type-I error rates in \% (nominal level $\alpha = 5\%$) for the parametric bootstrap (PBS) of the WTS and the MATS in the one-way layout for the log-normal distribution.}
	\label{app:logn}
	\begin{tabular}{c|c|c|c|c}
		\hline
		$d$ & Cov & $n$ & WTS (PBS)  & MATS (PBS)\\ 
		\hline
		\multirow{16}{*}{$d=4$} & \multirow{4}{*}{S1} &  (10, 10) & 2.0 &  3.5 \\ 
		&  & (10, 20) & 4.4 &  5.3 \\ 
		&  & (20, 10) & 4.3 &  5.5 \\ 
		&  & (20, 20) & 3.5 &  4.0 \\ \cline{2-5}
		& \multirow{4}{*}{S2} & (10, 10) & 1.9 &  3.2 \\ 
		&  & (10, 20) & 4.5 &  4.9 \\ 
		&  & (20, 10) & 4.4 &  5.4 \\ 
		&  & (20, 20) & 3.4 &  3.9 \\ \cline{2-5}
		& \multirow{4}{*}{S3} & (10, 10) & 6.8 &  6.1 \\ 
		&  & (10, 20) & 4.2 & 3.6 \\ 
		&  & (20, 10) & 14.9 &  13.8 \\ 
		&  & (20, 20) & 8.4 &  6.7 \\ \cline{2-5}
		& \multirow{4}{*}{S4} & (10, 10) & 7.0 &  6.2 \\ 
		&  & (10, 20) & 4.2 &  3.6 \\ 
		&  & (20, 10) & 15.4 &  13.5 \\ 
		&  & (20, 20) & 8.7 &  6.8 \\
		\hline \hline
		\multirow{16}{*}{$d=8$} & \multirow{4}{*}{S1}  & (10, 10) & 2.9 &  4.1 \\ 
		&  & (10, 20) & 3.9 &  5.6 \\ 
		& & (20, 10) & 4.2 &  5.3 \\ 
		& & (20, 20) & 3.0 & 4.0 \\  \cline{2-5}
		&  \multirow{4}{*}{S2} & (10, 10) & 2.8 &  2.5 \\ 
		& & (10, 20) & 4.0 &  4.8 \\ 
		& & (20, 10) & 4.1 &  4.1 \\ 
		& & (20, 20) & 3.0 &  3.1 \\  \cline{2-5}
		&  \multirow{4}{*}{S3} & (10, 10) & 6.3 & 6.0 \\ 
		& & (10, 20) & 3.5 &  3.6 \\ 
		& & (20, 10) & 15.0  & 12.7 \\ 
		& & (20, 20) & 7.7  & 6.7 \\  \cline{2-5}
		&  \multirow{4}{*}{S4} & (10, 10) & 7.0  & 6.0 \\ 
		& & (10, 20) & 3.6 &  2.8 \\ 
		& & (20, 10) & 15.5  & 13.6 \\ 
		& & (20, 20) & 8.0  & 7.0 \\ 
		\hline
	\end{tabular}
\end{table}

\begin{table}[H]
	\centering
	\caption{Type-I error rates in \% (nominal level $\alpha = 5\%$) for the parametric bootstrap (PBS) of the WTS and the MATS in the one-way layout for the $t_3$ distribution.}
	\label{app:t3}
	\begin{tabular}{c|c|c|c|c}
		\hline
		$d$ & Cov & $n$ & WTS (PBS) & MATS (PBS)\\ 
		\hline				
		\multirow{16}{*}{$d=4$} & \multirow{4}{*}{S1} & (10, 10) & 3.6 &  4.4 \\ 
		&  & (10, 20) & 4.8 & 4.8 \\ 
		&  & (20, 10) & 3.9 & 4.3 \\ 
		&  & (20, 20) & 4.0 &  4.5 \\ \cline{2-5}
		& \multirow{4}{*}{S2} & (10, 10) & 3.7 &  4.1 \\ 
		&  & (10, 20) & 4.7 &  4.6 \\ 
		&  & (20, 10) & 4.1 &  4.2 \\ 
		&  & (20, 20) & 4.0 & 4.3 \\ \cline{2-5}
		& \multirow{4}{*}{S3} & (10, 10) & 4.1 &  3.1 \\ 
		&  & (10, 20) & 4.3 &  4.2 \\ 
		&  & (20, 10) & 4.9 &  3.3 \\ 
		&  & (20, 20) & 3.9 &  3.8 \\ \cline{2-5}
		& \multirow{4}{*}{S4} & (10, 10) & 4.2 &  3.1 \\ 
		&  & (10, 20) & 4.4 &  3.9 \\ 
		&  & (20, 10) & 4.9 &  3.2 \\ 
		&  & (20, 20) & 4.0 &  4.0 \\ 
		\hline \hline
		\multirow{16}{*}{$d=8$} & \multirow{4}{*}{S1} & (10, 10) & 3.5 & 4.7 \\ 
		&  & (10, 20) & 4.8 &  4.7 \\ 
		&  & (20, 10) & 5.0 &  4.2 \\ 
		&  & (20, 20) & 3.9 &  4.7 \\ \cline{2-5}
		& \multirow{4}{*}{S2} & (10, 10) & 3.4  & 3.8 \\ 
		&  & (10, 20) & 4.8 &  4.1 \\ 
		&  & (20, 10) & 5.0 &  3.5 \\ 
		&  & (20, 20) & 3.9 &  3.9 \\ \cline{2-5}
		& \multirow{4}{*}{S3} & (10, 10) & 5.2  & 3.2 \\ 
		&  & (10, 20) & 3.5 &  4.2 \\ 
		&  & (20, 10) & 8.6 &  2.7 \\ 
		&  & (20, 20) & 4.0 & 3.6 \\ \cline{2-5}
		& \multirow{4}{*}{S4} & (10, 10) & 5.0 & 2.6 \\ 
		&  & (10, 20) & 3.6 &  3.6 \\ 
		&  & (20, 10) & 8.3 &  2.5 \\ 
		&  & (20, 20) & 3.8 &  3.2 \\ 
		\hline
	\end{tabular}
\end{table}

\begin{table}[H]
	\centering
	\caption{Type-I error rates in \% (nominal level $\alpha = 5\%$) for the parametric bootstrap (PBS) of the WTS and the MATS in the one-way layout for the double-exponential distribution.}
	\label{app:dexp}
	\begin{tabular}{c|c|c|c|c}
		\hline
		$d$ & Cov & $n$ & WTS (PBS) &   MATS (PBS)\\ 
		\hline
		\multirow{16}{*}{$d=4$} & \multirow{4}{*}{S1} & (10, 10) & 4.1  & 4.5 \\ 
		&  & (10, 20) & 5.0  &  4.8 \\ 
		&  & (20, 10) & 4.1 &  4.2 \\ 
		&  & (20, 20) & 5.1  & 5.6 \\ \cline{2-5}
		& \multirow{4}{*}{S2} & (10, 10) & 4.1 & 4.5 \\ 
		&  & (10, 20) & 4.9  & 4.8 \\ 
		&  & (20, 10) & 4.0  & 4.2 \\ 
		&  & (20, 20) & 5.1  & 5.4 \\ \cline{2-5}
		& \multirow{4}{*}{S3} & (10, 10) & 4.5  & 3.3 \\ 
		&  & (10, 20) & 4.3  & 4.4 \\ 
		&  & (20, 10) & 5.1  &  3.2 \\ 
		&  & (20, 20) & 4.7  & 4.9 \\ \cline{2-5}
		& \multirow{4}{*}{S4} & (10, 10) & 4.6 & 3.7 \\ 
		&  & (10, 20) & 4.6 &  4.3 \\ 
		&  & (20, 10) & 5.0  & 3.3 \\ 
		&  & (20, 20) & 4.7   & 4.7 \\ 
		\hline \hline
		\multirow{16}{*}{$d=8$} & \multirow{4}{*}{S1} & (10, 10) & 3.4 &  4.6 \\ 
		&  & (10, 20) & 5.2  & 4.8 \\ 
		&  & (20, 10) & 4.7 & 4.9 \\ 
		&  & (20, 20) & 4.3 & 4.7 \\ \cline{2-5}
		& \multirow{4}{*}{S2} & (10, 10) & 3.6  & 3.4 \\ 
		&  & (10, 20) & 5.3  & 4.3 \\ 
		&  & (20, 10) & 4.6  & 4.3 \\ 
		&  & (20, 20) & 4.3  & 4.3 \\ \cline{2-5}
		& \multirow{4}{*}{S3} & (10, 10) & 4.8   & 3.0 \\ 
		&  & (10, 20) & 4.1  & 4.2 \\ 
		&  & (20, 10) & 8.1  & 3.0 \\ 
		&  & (20, 20) & 5.0 & 3.6 \\ \cline{2-5}
		& \multirow{4}{*}{S4} & (10, 10) & 4.7  & 2.5 \\ 
		&  & (10, 20) & 4.2 & 3.7 \\ 
		&  & (20, 10) & 7.9 & 2.5 \\ 
		&  & (20, 20) & 4.9 & 3.5 \\ 
		\hline
	\end{tabular}
\end{table}

\begin{table}[H]
	\centering
	\caption{Type-I error rates in \% (nominal level $\alpha = 5\%$) for the {\color{black} WTS with $\chi^2$-approximation and parametric bootstrap (PBS) and the MATS with wild bootstrap (wild), parametric bootstrap (PBS) and nonparametric bootstrap (NPBS)} when testing for an effect of factor $A$ in a two-way layout with $d=4$ dimensional observations.}
	\label{app:facA}
	\begin{tabular}{c|c|c|cc|ccc}
		\hline
		distr & Cov & n & WTS ($\chi^2$) & WTS (PBS) & MATS (wild) & MATS (PBS) & MATS (NPBS) \\ 
		\hline
		\multirow{16}{*}{normal} & \multirow{4}{*}{S8} &  $\vn^{(1)}$& 10.4 & 4.7 & 6.1 & 4.6 & 4.4 \\ 
		&  & $\vn^{(2)}$ & 9.5 & 4.5 & 6.4 & 5.1 & 4.9 \\ 
		& & $\vn^{(3)}$ & 11.2 & 5.4 & 6.6 & 5.3 & 4.8 \\ 
		&  & $\vn^{(4)}$ &  6.8&     5.0    &   5.6   &   5.3    &   5.2  \\ \cline{2-8}
		& \multirow{4}{*}{S9} & $\vn^{(1)}$ & 10.4 & 4.8 & 6.2 & 4.7 & 4.4 \\ 
		&  & $\vn^{(2)}$ & 9.5 & 4.4 & 6.6 & 5.3 & 4.9 \\ 
		&  & $\vn^{(3)}$ & 11.2 & 5.5 & 6.5 & 4.9 & 4.6 \\ 
		&  & $\vn^{(4)}$ & 6.8 & 4.9 & 5.7 & 5.1 & 5.1 \\ \cline{2-8}
		& \multirow{4}{*}{S10} & $\vn^{(1)}$ & 9 & 4.7 & 5.8 & 4.5 & 4.2 \\ 
		&  & $\vn^{(2)}$ & 11.1 & 5.2 & 6.4 & 4.8 & 4.3 \\ 
		&  & $\vn^{(3)}$ & 14.8 & 6.1 & 7.3 & 4.9 & 4 \\ 
		&  & $\vn^{(4)}$ & 7.4 & 5 & 5.5 & 4.7 & 4.7 \\ \cline{2-8}
		& \multirow{4}{*}{S11} & $\vn^{(1)}$ & 9.2 & 4.7 & 5.8 & 4.4 & 4.1 \\ 
		&  & $\vn^{(2)}$ & 10.2 & 4.9 & 6 & 4.4 & 4.1 \\ 
		&  & $\vn^{(3)}$ & 13.3 & 6 & 7 & 4.8 & 4.2 \\ 
		&  & $\vn^{(4)}$ & 7 & 5.1 & 5.4 & 4.6 & 4.5 \\ 
		\hline\hline
		\multirow{16}{*}{$\chi^2_3$} & \multirow{4}{*}{S8}  & $\vn^{(1)}$ & 9.7 & 4.6 & 6.7 & 5.1 & 4.5 \\ 
		&  & $\vn^{(2)}$ & 9.2 & 3.9 & 6.7 & 5.1 & 4.5 \\ 
		&  & $\vn^{(3)}$ & 10.7 & 4.7 & 6.7 & 5 & 4.4 \\ 
		&  & $\vn^{(4)}$ & 6.6 & 4.3 & 4.7 & 4.3 & 4.1 \\ \cline{2-8}
		& \multirow{4}{*}{S9}  & $\vn^{(1)}$ & 9.7 & 4.6 & 6.5 & 4.7 & 4.1 \\ 
		&  & $\vn^{(2)}$ & 9.2 & 3.9 & 7 & 5.1 & 4.3 \\ 
		&  & $\vn^{(3)}$ & 10.7 & 4.8 & 6.6 & 5 & 4.2 \\ 
		&  & $\vn^{(4)}$ & 6.6 & 4.5 & 4.9 & 4.4 & 4.2 \\ \cline{2-8}
		& \multirow{4}{*}{S10}  & $\vn^{(1)}$ & 8.8 & 4.5 & 6.1 & 4.3 & 3.6 \\ 
		&  & $\vn^{(2)}$ & 11.2 & 5 & 7.6 & 5.2 & 4.1 \\ 
		&  & $\vn^{(3)}$ & 16 & 7 & 9.4 & 6.5 & 5.3 \\ 
		&  & $\vn^{(4)}$ & 7.7 & 5.5 & 5.6 & 4.9 & 4.5 \\ \cline{2-8}
		& \multirow{4}{*}{S11}  & $\vn^{(1)}$ & 8.5 & 4.2 & 5.6 & 3.7 & 3 \\ 
		&  & $\vn^{(2)}$ & 9.9 & 4.3 & 6.7 & 4.7 & 3.8 \\ 
		&  & $\vn^{(3)}$ & 13.9 & 6.4 & 8.7 & 5.9 & 4.4 \\ 
		&  & $\vn^{(4)}$ & 7.2 & 5 & 5.6 & 4.9 & 4.3 \\ 
		
		\hline
	\end{tabular}
\end{table}

\begin{table}[H]
	\centering
	\caption{Type-I error rates in \% (nominal level $\alpha = 5\%$) for the {\color{black} WTS with $\chi^2$-approximation and parametric bootstrap (PBS) and the MATS with wild bootstrap (wild), parametric bootstrap (PBS) and nonparametric bootstrap (NPBS) }when testing for an effect of factor $B$ in a two-way layout with $d=4$ dimensional observations.}
	\label{app:facB}
	\begin{tabular}{c|c|c|cc|ccc}
		\hline
		distr & Cov & n & WTS($\chi^2$) & WTS (PBS) &  MATS (wild) & MATS (PBS) & MATS (NPBS)\\ 
		\hline
		\multirow{16}{*}{normal} & \multirow{4}{*}{S8} & $\vn^{(1)}$ &  10.7 & 5.1 & 6.9 & 5.3 & 4.8 \\ 
		&  & $\vn^{(2)}$ & 9.2 & 4.6 & 6 & 4.8 & 4.5 \\ 
		&  & $\vn^{(3)}$ & 10.2 & 4.6 & 6.3 & 5 & 4.7 \\ 
		&  & $\vn^{(4)}$ & 6.5 &    4.4   &    5.0    &  4.7    &   4.7  \\ \cline{2-8}
		& \multirow{4}{*}{S9} & $\vn^{(1)}$ &   10.7 & 5.2 & 6.6 & 5.1 & 4.8 \\ 
		&  & $\vn^{(2)}$ & 9.2 & 4.7 & 6.1 & 4.9 & 4.8 \\ 
		&  & $\vn^{(3)}$ & 10.2 & 4.5 & 6.2 & 4.5 & 4.3 \\ 
		&  & $\vn^{(4)}$ & 6.5 & 4.5 & 5 & 4.5 & 4.4 \\ \cline{2-8}
		&\multirow{4}{*}{S10} & $\vn^{(1)}$ &   8.7 & 4.9 & 5.8 & 4.7 & 4.4 \\ 
		&  & $\vn^{(2)}$ & 10.3 & 4.6 & 5.9 & 4.5 & 4.1 \\ 
		&  & $\vn^{(3)}$ & 13.4 & 5.3 & 6.4 & 4 & 3.3 \\ 
		&  & $\vn^{(4)}$ & 6.9 & 4.6 & 5.2 & 4.7 & 4.6 \\ \cline{2-8}
		& \multirow{4}{*}{S11} & $\vn^{(1)}$ &  9 & 4.9 & 5.7 & 4.4 & 4.3 \\ 
		&  & $\vn^{(2)}$ & 9.4 & 4.7 & 6.1 & 4.4 & 4.1 \\ 
		&  & $\vn^{(3)}$ & 12.4 & 5.1 & 6.4 & 4.1 & 3.4 \\ 
		&  & $\vn^{(4)}$ & 6.4 & 4.6 & 5.2 & 4.4 & 4.3 \\ 
		\hline\hline
		\multirow{16}{*}{$\chi^2_3$} & \multirow{4}{*}{S8}  & $\vn^{(1)}$ & 10.3 & 4.9 & 6.4 & 4.9 & 4.3 \\ 
		&  & $\vn^{(2)}$ & 9.4 & 4 & 6.2 & 4.8 & 4.3 \\ 
		&  & $\vn^{(3)}$ & 10.7 & 4.5 & 5.9 & 4.7 & 4.1 \\ 
		&  & $\vn^{(4)}$ & 7.2 & 4.7 & 5.1 & 4.7 & 4.5 \\ \cline{2-8}
		& \multirow{4}{*}{S9} & $\vn^{(1)}$ &   10.3 & 4.9 & 6.8 & 4.8 & 4.2 \\ 
		&  & $\vn^{(2)}$ & 9.4 & 4 & 6.3 & 4.6 & 4 \\ 
		&  & $\vn^{(3)}$ & 10.7 & 4.5 & 6 & 4.4 & 4 \\ 
		&  & $\vn^{(4)}$ & 7.2 & 4.9 & 5.4 & 5 & 4.8 \\ \cline{2-8}
		& \multirow{4}{*}{S10} & $\vn^{(1)}$ &   9 & 4.5 & 6.1 & 4.4 & 3.8 \\ 
		&  & $\vn^{(2)}$ & 10.6 & 4.5 & 6.5 & 4.4 & 3.2 \\ 
		&  & $\vn^{(3)}$ & 13.7 & 5.4 & 7 & 4.4 & 3.3 \\ 
		&  & $\vn^{(4)}$ & 7.7 & 5.1 & 6 & 5.1 & 4.6 \\ \cline{2-8}
		& \multirow{4}{*}{S11} & $\vn^{(1)}$ &   9.2 & 4.6 & 6.2 & 4.4 & 3.7 \\ 
		&  & $\vn^{(2)}$ & 9.7 & 4.2 & 6.3 & 4.1 & 2.9 \\ 
		&  & $\vn^{(3)}$ & 12.4 & 5.1 & 6.4 & 4.2 & 3.1 \\ 
		&  & $\vn^{(4)}$ & 7.3 & 5.2 & 6 & 5.1 & 4.3 \\ 
		\hline
	\end{tabular}
\end{table}

\begin{table}[H]
	\centering
	\caption{Type-I error rates in \% (nominal level $\alpha = 5\%$) for the {\color{black} WTS with $\chi^2$-approximation and parametric bootstrap (PBS) and the MATS with wild bootstrap (wild), parametric bootstrap (PBS) and nonparametric bootstrap (NPBS) }when testing the interaction hypothesis in a two-way layout with $d=4$ dimensional observations.}
	\label{app:interaction}
	\begin{tabular}{c|c|c|cc|ccc}
		\hline
		distr & Cov & n & WTS($\chi^2$) & WTS (PBS) &  MATS (wild) & MATS (PBS) & MATS(NPBS)\\ 
		\hline
		\multirow{16}{*}{normal} & \multirow{4}{*}{S8} & $\vn^{(1)}$ & 10.8 & 4.8 & 6.9 & 5.4 & 5 \\ 
		&  & $\vn^{(2)}$ & 10 & 4.7 & 6.5 & 5.4 & 5.2 \\ 
		&  & $\vn^{(3)}$ & 10 & 4.7 & 6.8 & 5.3 & 5 \\ 
		&  & $\vn^{(4)}$ &  6.3   &  4.3     &  5.0    &  4.6     &  4.6  \\ \cline{2-8}
		& \multirow{4}{*}{S9} & $\vn^{(1)}$ & 10.8 & 4.9 & 6.9 & 5.1 & 4.9 \\ 
		&  & $\vn^{(2)}$ & 10 & 4.8 & 6.3 & 5.1 & 4.8 \\ 
		&  & $\vn^{(3)}$ & 10 & 4.8 & 6.4 & 5 & 4.8 \\ 
		&  & $\vn^{(4)}$ & 6.3 & 4.3 & 5 & 4.7 & 4.4 \\\cline{2-8} 
		& \multirow{4}{*}{S10} & $\vn^{(1)}$ & 8.8 & 5.1 & 6.2 & 4.8 & 4.5 \\ 
		&  & $\vn^{(2)}$ & 10.9 & 4.8 & 6.6 & 4.9 & 4.5 \\ 
		&  & $\vn^{(3)}$ & 13.8 & 5.7 & 7.5 & 4.9 & 4.1 \\ 
		&  & $\vn^{(4)}$ & 6.9 & 4.5 & 5 & 4.3 & 4.2 \\ \cline{2-8}
		& \multirow{4}{*}{S11} & $\vn^{(1)}$ & 9.1 & 4.9 & 6.4 & 4.8 & 4.4 \\ 
		&  & $\vn^{(2)}$ & 10.1 & 4.8 & 6.6 & 4.7 & 4.5 \\ 
		&  & $\vn^{(3)}$ & 12.9 & 5.4 & 7.3 & 4.7 & 4.1 \\ 
		&  & $\vn^{(4)}$ & 6.5 & 4.5 & 4.9 & 4.4 & 4.2 \\ 
		\hline\hline
		\multirow{16}{*}{$\chi^2_3$} & \multirow{4}{*}{S8} & $\vn^{(1)}$ & 9.5 & 4.3 & 6.3 & 4.8 & 4.3 \\ 
		&  & $\vn^{(2)}$ & 9.4 & 4.3 & 6.3 & 4.8 & 4.4 \\ 
		&  & $\vn^{(3)}$ & 9.8 & 4.1 & 5.9 & 4.6 & 4 \\ 
		&  & $\vn^{(4)}$ & 6.6 & 4.6 & 4.8 & 4.4 & 4.2 \\ \cline{2-8}
		& \multirow{4}{*}{S9} & $\vn^{(1)}$ & 9.5 & 4.3 & 6.3 & 4.6 & 4.1 \\ 
		&  & $\vn^{(2)}$ & 9.4 & 4.1 & 6.5 & 4.8 & 4.2 \\ 
		&  & $\vn^{(3)}$ & 9.8 & 4.2 & 5.9 & 4.6 & 4 \\ 
		&  & $\vn^{(4)}$ & 6.6 & 4.6 & 5.1 & 4.6 & 4.4 \\ \cline{2-8}
		& \multirow{4}{*}{S10} & $\vn^{(1)}$ & 7.9 & 4.1 & 5.9 & 4.3 & 3.6 \\ 
		&  & $\vn^{(2)}$ & 10.3 & 4.2 & 5.9 & 3.9 & 2.9 \\ 
		&  & $\vn^{(3)}$ & 13 & 4.3 & 6.4 & 3.7 & 2.4 \\ 
		&  & $\vn^{(4)}$ & 6.8 & 4.7 & 5.5 & 4.6 & 4 \\ \cline{2-8}
		& \multirow{4}{*}{S11} & $\vn^{(1)}$ & 8.2 & 4 & 5.8 & 4 & 3.4 \\ 
		&  & $\vn^{(2)}$ & 9.7 & 4.2 & 6 & 3.9 & 2.9 \\ 
		&  & $\vn^{(3)}$ & 11.7 & 4.3 & 6.2 & 3.5 & 2.4 \\ 
		&  & $\vn^{(4)}$ & 6.7 & 4.8 & 5.4 & 4.4 & 4.2 \\ 
		\hline
	\end{tabular}
\end{table}

\section{Another data example}\label{datex2}

\begin{figure}
	\centering
	\includegraphics[width = 0.8\textwidth]{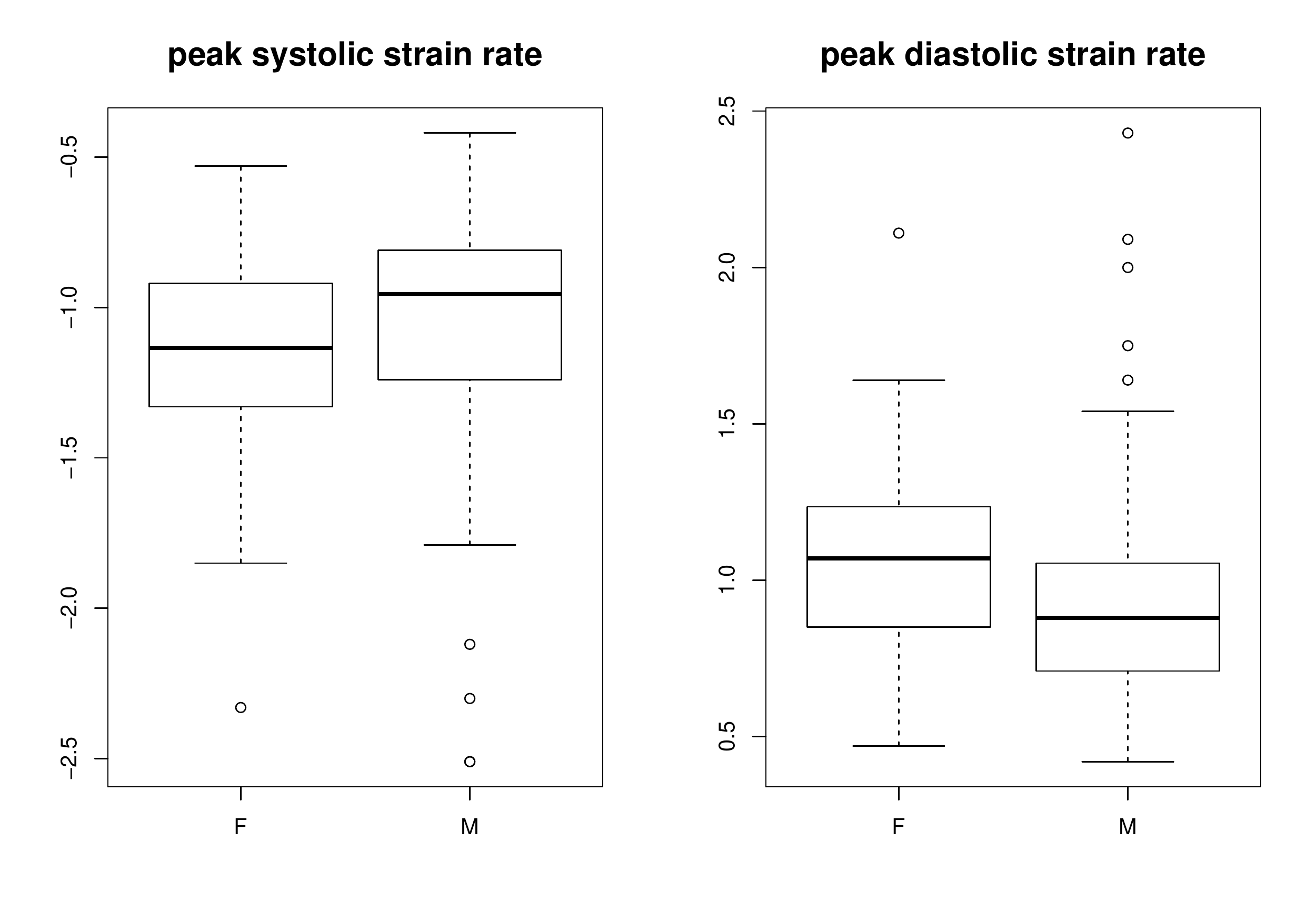}
	\caption{Boxplots of the systolic and diastolic peak strain rate for female and male patients.}
	\label{fig:strainrate}
\end{figure}

\begin{figure}
	\centering
	\includegraphics[width = \textwidth]{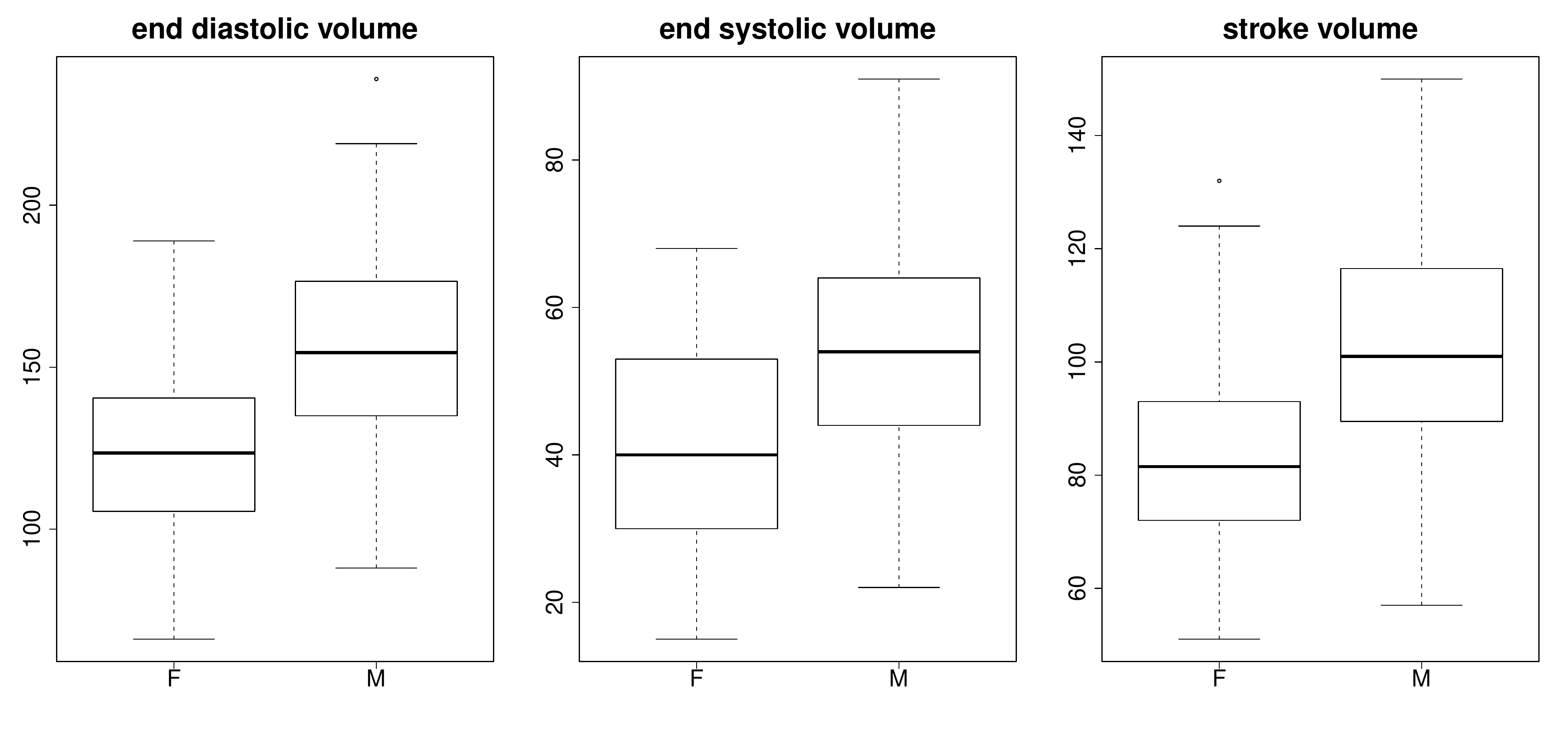}
	\caption{Boxplots of the end diastolic and systolic volume as well as stroke volume for female and male patients.}
	\label{fig:volume}
\end{figure}

As {\color{black}our second} data example, we consider cardiological measurements in the left ventricle of 188 healthy patients, that were recorded at the University clinic Ulm, Germany.
Variables of interest are the peak systolic and diastolic strain rate (PSSR and PDSR, respectively), measured in 
circumferential direction, the end systolic and diastolic volume (ESV and EDV, respectively) as well as the stroke volume (SV). The empirical covariance matrix is singular in this example, since stroke volume is calculated as the difference between end diastolic volume and end systolic volume. 
The empirical covariance matrices can be found {\color{black} in Section \ref{covmat} below}. {\color{black} Note that this data example is somewhat artificial, since the reason for the singularity of the empirical covariance matrix is known and one would usually drop one of the three variables involved in the collinearity.}
We consider a one-way layout analyzing the factor 'gender' (female vs.~male). 
Some descriptive statistics of the measurements for this factor are displayed in Table \ref{tab:descriptive}. 
Boxplots of the systolic and diastolic measurements are in Figures \ref{fig:strainrate} and \ref{fig:volume}, respectively.

\begin{table}[H]
	\centering
	\small
	\caption{Descriptive statistics of the cardiology data. Volume measurements (EDV, ESV and SV) are in $m\ell$, peak strain rate measurements are in $1/sec.$}
	\label{tab:descriptive}
	\begin{tabular}{c|c||ccccc||ccccc}
		Gender & $n$ & \multicolumn{5}{c||}{Mean}& \multicolumn{5}{c}{Sd}\\\hline
		& & EDV  &  ESV  &   SV & PSSR & PDSR & EDV  &  ESV  &   SV & PSSR & PDSR\\
		\hline
		female & 92 & 124.37  &  41.39 &   82.98  &  -1.16  &   1.07 & 25.25 &   13.38   & 15.77   &  0.32 &    0.30\\
		male & 96 & 157.02  &  54.98  & 102.04  &  -1.07  &   0.94 &  31.02   & 15.98  &  18.70  &   0.39   &  0.35\\
	\end{tabular}
\end{table}

We now want to analyze, whether there is a significant difference in the multivariate means for female and male patients. The null hypothesis of interest thus is $H_0^{(1)}: \{(\vP_2 \kp \vI_5) \vmu = \vnull\}$.
Since the covariance matrix is singular in this example, we cannot apply the Wald-type test. Thus, we consider the parametric bootstrap approach of the MATS which yielded the best results in the simulation study.
Computation of the MATS results in a value of $Q_N = 171.0011$ and the parametric bootstrap routine with 10,000 bootstrap runs gives a $p$-value of $p < 0.0001$, implying that there is indeed a significant effect of gender on the measurements.

\begin{figure}[h]
	\centering
	\includegraphics[width = 0.8\textwidth]{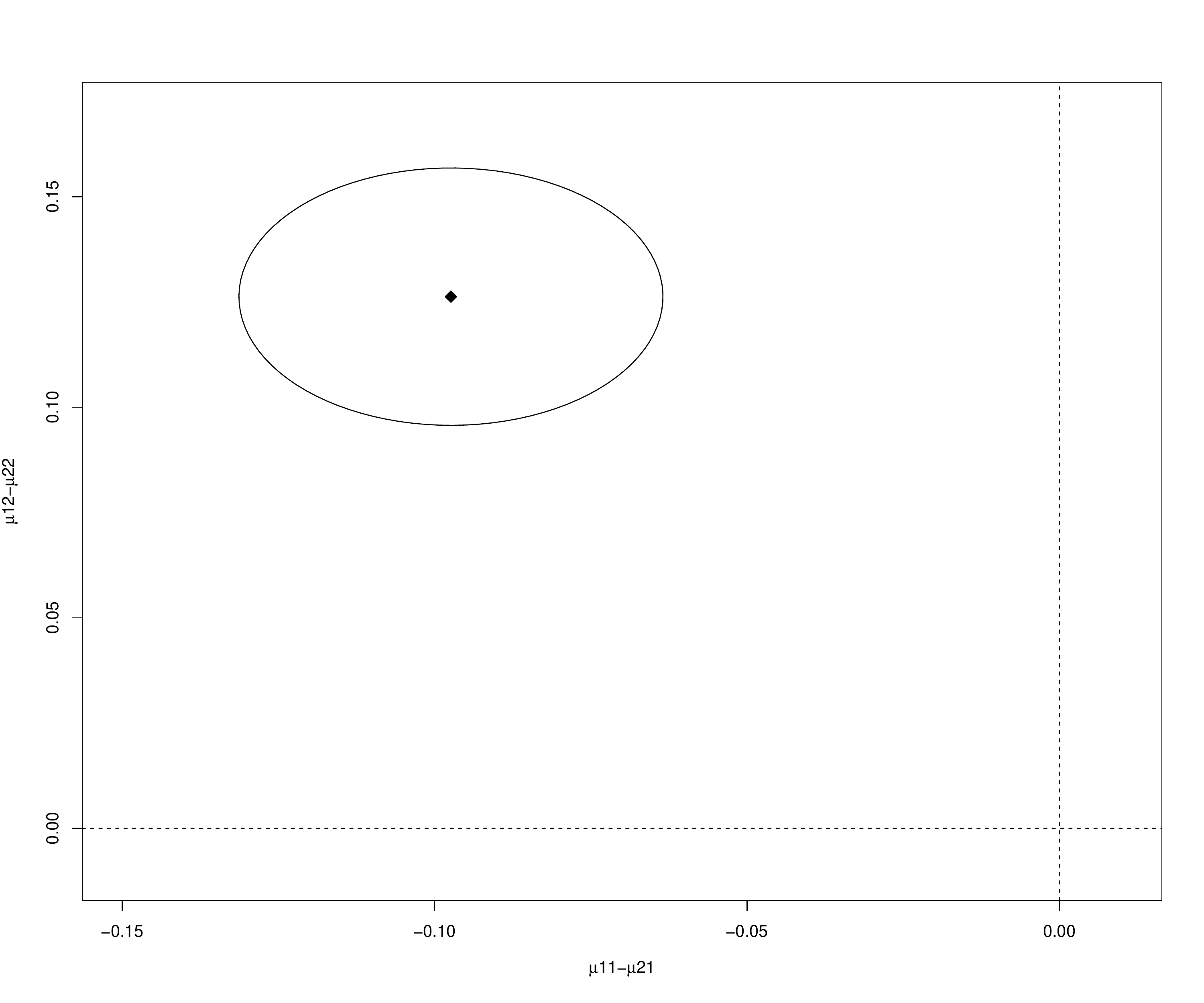}
	\caption{95 \% confidence ellipse for $\vmu_1 - \vmu_2$. The $\scriptstyle \blacklozenge$ denotes the center $\vT\volX_{\bullet}$ of the ellipse. Female patients seem to have a lower systolic peak strain rate but a higher diastolic peak strain rate than male patients.}
	\label{confellipse}
\end{figure}

In a second step we want to derive a confidence region for the factor 'Gender'. Here, we restrict our analyses to the strain rate measurements in order to be able to plot the confidence ellipsoids for the contrast of interest.
That is, we consider the null hypothesis $H_0^{(2)}: \{\vT\vmu = \vnull \} =  \{\mu_{11} - \mu_{21} = \mu_{12}-\mu_{22} \} = \{\vmu_1 -\vmu_2 = \vnull\}$, where $\mu_{ij}$ is the corresponding mean value of measurement $j$ (systolic vs.~diastolic measurement) in group $i$ (female vs.~male) and $$\vT = \left(\begin{array}{cccc}
1 & 0 & -1 & 0\\
0 & 1 & 0 & -1\\
\end{array}\right).$$
The parametric bootstrap procedure with 10,000 bootstrap runs leads to a $p$-value of $p = 0.0146$ for the MATS, i.e., there is a significant effect of gender on the peak strain rate. We can now construct a confidence ellipsoid as described in Section \ref{confint} based on the parametric bootstrap quantile $c_{1-\alpha}^*$. Therefore, we need to compute the eigenvalues $\lambda_j$ and eigenvectors $\ve_j, j = 1, 2$ of $\vT \vwhD_N \vT$. The ellipse is centered at $\vT \volX_{\bullet} = (-0.097,  0.126)^\top$. For the eigendecomposition of $\vT \vwhD_N \vT$, we obtain $\vlam = (0.508,  0.412)$ as well as $\ve_1 = (-1, 0)^\top$ and $\ve_2 =(0, -1)^\top$, that is, the confidence ellipse extends $\sqrt{\lambda_1 \cdot c_{1-\alpha}^*/N} = 0.013$ units in the direction of $\ve_1$ and  0.012 units in the direction of $\ve_2$. The corresponding ellipse is displayed in Figure \ref{confellipse}. It turns out that female patients have a lower systolic peak strain rate but a higher diastolic peak strain rate than male patients, a finding that is confirmed by the descriptive analysis in Table \ref{tab:descriptive} and Figure \ref{fig:strainrate}.

\subsection{Covariance Matrices}\label{covmat}

For convenience, we included the estimated covariance matrices of the data example for female and male patients and the five outcome variables end diastolic volume (EDV), end systolic volume (ESV), stroke volume (SV), peak systolic strain rate (PSSR) and peak diastolic strain rate (PDSR) in this section. 
Covariance matrices, rounded to three decimals, for female patients (EDV, ESV, SV, PSSR, PDSR):
$$
\left( \begin{array}{ccccc}     
6.931 & 3.087 & 3.844  &  0.022 &  -0.012\\
3.087 & 1.947 & 1.140  &  0.018  & -0.013\\
3.844 & 1.140 & 2.704   & 0.004   & 0.002\\
0.022 & 0.018 & 0.004  &  0.001  & -0.001\\
-0.012 & -0.013 & 0.002  & -0.001  &  0.001\\
\end{array} \right)
$$
and male patients:
$$
\left( \begin{array}{ccccc}     
10.025 & 4.522 & 5.503 &   0.043 &  -0.021\\
4.522 & 2.661 & 1.862  &  0.027 &  -0.018\\
5.503 & 1.862 & 3.641 &   0.016 &  -0.003\\
0.043 & 0.027 & 0.016 &   0.002 &  -0.001\\
-0.021 & -0.018 & -0.003  & -0.001 &   0.001\\
\end{array} \right).
$$

%

{\color{black}
	\subsection{Problem with the ATS}\label{probATS}
	
	Finally, we want to demonstrate the problems that can arise when using the ATS $\tilde{Q}_N$ in multivariate data. Since the asymptotic distribution of the ATS depends on unknown parameters \citep{brunner:2001,friedrich2017permuting}, it is approximated by an $\mathcal{F}$-distribution. In particular, the scaled statistic
	$$
	F_N = \frac{N}{\tr(\vT \vSigmah)} \volX_{\bullet}^\top \vT \volX_{\bullet}
	$$
	is approximated by an $\mathcal{F}(\hnu, \infty)$-distribution with $\hnu = \tr^2(\vT \vwhSigma)/\tr(\vT \vwhSigma)^2$ degrees of freedom. We consider only the peak strain rate measurements. The ATS then results in a test statistic of $F_N = 5.2$, while the corresponding quantile of the $\mathcal{F}$-distribution is 3.51, resulting in a $p$-value of 0.02. If we now change the units of the peak systolic strain rate from $1/sec$ to $1/min$, the test statistic becomes $F_N = 3.51$ with an $\mathcal{F}$-quantile of 3.84, resulting in a $p$-value of 0.06. By changing the units in one component, we have therefore changed the significance of the outcome at 5\% level.
	Thus, the ATS should only be applied if observations are measured on the same scale as in repeated measurements but not to multivariate data in general. 
	
}

\bibliography{Literatur}{}

\begin{thebibliography}{10}

\bibitem{bartlett}
M.~S. Bartlett.
\newblock A note on tests of significance in multivariate analysis.
\newblock {\em Mathematical Proceedings of the Cambridge Philosophical
  Society}, 35(02):180--185, 1939.
\newblock Cambridge University Press.

\bibitem{bathke2008compare}
A.~C. Bathke, S.~W. Harrar, and L.~V. Madden.
\newblock How to compare small multivariate samples using nonparametric tests.
\newblock {\em Computational Statistics \& Data Analysis}, 52(11):4951--4965,
  2008.

\bibitem{Beyersmann2013}
J.~Beyersmann, S.~D. Termini, and M.~Pauly.
\newblock Weak convergence of the wild bootstrap for the {A}alen--{J}ohansen
  estimator of the cumulative incidence function of a competing risk.
\newblock {\em Scandinavian Journal of Statistics}, 40(3):387--402, 2013.

\bibitem{bickel1981some}
P.~J. Bickel and D.~A. Freedman.
\newblock Some asymptotic theory for the bootstrap.
\newblock {\em The Annals of Statistics}, pages 1196--1217, 1981.

\bibitem{brunner:2001}
E.~Brunner.
\newblock Asymptotic and approximate analysis of repeated measures designs
  under heteroscedasticity.
\newblock {\em Mathematical Statistics with Applications in Biometry}, 2001.

\bibitem{brunner2016rank}
E.~Brunner, F.~Konietschke, M.~Pauly, and M.~L. Puri.
\newblock Rank-based procedures in factorial designs: hypotheses about
  non-parametric treatment effects.
\newblock {\em Journal of the Royal Statistical Society: Series B (Statistical
  Methodology)}, 2016.

\bibitem{brunner:99}
E.~Brunner, U.~Munzel, and M.~L. Puri.
\newblock Rank-score tests in factorial designs with repeated measures.
\newblock {\em Journal of Multivariate Analysis}, 70(2):286--317, 1999.

\bibitem{brunnerPuri}
E.~Brunner and M.~L. Puri.
\newblock Nonparametric methods in factorial designs.
\newblock {\em Statistical papers}, 42(1):1--52, 2001.

\bibitem{cameron2008bootstrap}
A.~C. Cameron, J.~B. Gelbach, and D.~L. Miller.
\newblock Bootstrap-based improvements for inference with clustered errors.
\newblock {\em The Review of Economics and Statistics}, 90(3):414--427, 2008.

\bibitem{cameron2015practitioner}
A.~C. Cameron and D.~L. Miller.
\newblock A practitioner's guide to cluster-robust inference.
\newblock {\em Journal of Human Resources}, 50(2):317--372, 2015.

\bibitem{ChungRomano:2013}
E.~Chung and J.~P. Romano.
\newblock Multivariate and multiple permutation tests.
\newblock {\em Journal of Econometrics}, 193(1):76--91, 2016.

\bibitem{csorgo1992law}
S.~Cs{\"o}rgo.
\newblock On the law of large numbers for the bootstrap mean.
\newblock {\em Statistics \& probability letters}, 14(1):1--7, 1992.

\bibitem{davidson2008wild}
R.~Davidson and E.~Flachaire.
\newblock The wild bootstrap, tamed at last.
\newblock {\em Journal of Econometrics}, 146(1):162--169, 2008.

\bibitem{Dempster1958}
A.~P. Dempster.
\newblock A high dimensional two sample significance test.
\newblock {\em The Annals of Mathematical Statistics}, 29(4):995--1010, 1958.

\bibitem{Dempster1960}
A.~P. Dempster.
\newblock A significance test for the separation of two highly multivariate
  small samples.
\newblock {\em Biometrics}, 16(1):41--50, 1960.

\bibitem{friedrich2017permuting}
S.~Friedrich, E.~Brunner, and M.~Pauly.
\newblock Permuting longitudinal data in spite of the dependencies.
\newblock {\em Journal of Multivariate Analysis}, 153:255--265, 2017.

\bibitem{Friedrich2016}
S.~Friedrich, F.~Konietschke, and M.~Pauly.
\newblock A wild bootstrap approach for nonparametric repeated measurements.
\newblock {\em Computational Statistics \& Data Analysis}, 2016.

\bibitem{harrar2012modified}
S.~W. Harrar and A.~C. Bathke.
\newblock A modified two-factor multivariate analysis of variance: asymptotics
  and small sample approximations.
\newblock {\em Annals of the Institute of Statistical Mathematics},
  64(1):135--165, 2012.

\bibitem{hasler2008multiple}
M.~Hasler and L.~A. Hothorn.
\newblock Multiple contrast tests in the presence of heteroscedasticity.
\newblock {\em Biometrical Journal}, 50(5):793--800, 2008.

\bibitem{Hotelling:1951}
H.~Hotelling.
\newblock A generalized $t$-test and measure of multivariate dispersion.
\newblock In {\em Proceedings of the Second Berkeley Symposium on Mathematical
  Statistics and Probability}. The Regents of the University of California,
  1951.

\bibitem{hothorn2008simultaneous}
T.~Hothorn, F.~Bretz, and P.~Westfall.
\newblock Simultaneous inference in general parametric models.
\newblock {\em Biometrical Journal}, 50(3):346--363, 2008.

\bibitem{johnson}
R.~A. Johnson and D.~W. Wichern.
\newblock {\em Applied multivariate statistical analysis}.
\newblock 6th edition, Prentice Hall, 2007.

\bibitem{Kon:2015}
F.~Konietschke, A.~Bathke, S.~Harrar, and M.~Pauly.
\newblock Parametric and nonparametric bootstrap methods for general {MANOVA}.
\newblock {\em Journal of Multivariate Analysis}, 140:291--301, 2015.

\bibitem{krishnamoorthy2010}
K.~Krishnamoorthy and F.~Lu.
\newblock A parametric bootstrap solution to the {MANOVA} under
  heteroscedasticity.
\newblock {\em Journal of Statistical Computation and Simulation},
  80(8):873--887, 2010.

\bibitem{Lawley}
D.~Lawley.
\newblock A generalization of fisher's z test.
\newblock {\em Biometrika}, 30(1-2):180--187, 1938.

\bibitem{lin1997non}
D.~Lin.
\newblock Non-parametric inference for cumulative incidence functions in
  competing risks studies.
\newblock {\em Statistics in Medicine}, 16(8):901--910, 1997.

\bibitem{liu2011nonparametric}
C.~Liu, A.~C. Bathke, and S.~W. Harrar.
\newblock A nonparametric version of wilks' lambda - asymptotic results and
  small sample approximations.
\newblock {\em Statistics \& Probability Letters}, 81(10):1502--1506, 2011.

\bibitem{liu1988bootstrap}
R.~Y. Liu.
\newblock Bootstrap procedures under some non-iid models.
\newblock {\em The Annals of Statistics}, 16(4):1696--1708, 1988.

\bibitem{mammen93}
E.~Mammen.
\newblock {\em When does bootstrap work? {A}symptotic results and simulations}.
\newblock Springer Science \& Business Media, 1993.

\bibitem{marcus1976closed}
R.~Marcus, P.~Eric, and K.~R. Gabriel.
\newblock On closed testing procedures with special reference to ordered
  analysis of variance.
\newblock {\em Biometrika}, 63(3):655--660, 1976.

\bibitem{PEB}
M.~Pauly, D.~Ellenberger, and E.~Brunner.
\newblock Analysis of high-dimensional one group repeated measures designs.
\newblock {\em Statistics}, 49:1243--1261, 2015.

\bibitem{pesarin-salmaso-book}
F.~Pesarin and L.~Salmaso.
\newblock {\em Permutation tests for complex data: theory, applications and
  software}.
\newblock John Wiley \& Sons, 2010.

\bibitem{Pesarin:2012}
F.~Pesarin and L.~Salmaso.
\newblock A review and some new results on permutation testing for multivariate
  problems.
\newblock {\em Statistics and Computing}, 22(2):639--646, 2012.

\bibitem{Pillai}
K.~Pillai.
\newblock Some new test criteria in multivariate analysis.
\newblock {\em The Annals of Mathematical Statistics}, 26(1):117--121, 1955.

\bibitem{R}
{R Core Team}.
\newblock {\em R: A Language and Environment for Statistical Computing}.
\newblock R Foundation for Statistical Computing, Vienna, Austria, 2016.

\bibitem{Rao}
C.~Rao and S.~Mitra.
\newblock {\em Generalized inverse of matrices and its applications}.
\newblock Wiley New York, 1971.

\bibitem{smaga2016bootstrap}
{\L}.~Smaga.
\newblock Bootstrap methods for multivariate hypothesis testing.
\newblock {\em Communications in Statistics-Simulation and Computation}, 2016.
\newblock Just accepted.

\bibitem{sonnemann2008general}
E.~Sonnemann.
\newblock General solutions to multiple testing problems.
\newblock {\em Biometrical Journal}, 50(5):641--656, 2008.

\bibitem{srivastava2013tests}
M.~S. Srivastava and T.~Kubokawa.
\newblock Tests for multivariate analysis of variance in high dimension under
  non-normality.
\newblock {\em Journal of Multivariate Analysis}, 115:204--216, 2013.

\bibitem{VallejoAto2012}
G.~Vallejo and M.~Ato.
\newblock Robust tests for multivariate factorial designs under
  heteroscedasticity.
\newblock {\em Behavior research methods}, 44(2):471--489, 2012.

\bibitem{Vallejo}
G.~Vallejo, M.~Fern{\'a}ndez, and P.~E. Livacic-Rojas.
\newblock Analysis of unbalanced factorial designs with heteroscedastic data.
\newblock {\em Journal of Statistical Computation and Simulation},
  80(1):75--88, 2010.

\bibitem{van2011robust}
S.~Van~Aelst and G.~Willems.
\newblock Robust and efficient one-way {MANOVA} tests.
\newblock {\em Journal of the American Statistical Association},
  106(494):706--718, 2011.

\bibitem{van2013fast}
S.~Van~Aelst and G.~Willems.
\newblock Fast and robust bootstrap for multivariate inference: the {R package
  FRB}.
\newblock {\em Journal of Statistical Software}, 53(3):1--32, 2013.

\bibitem{Wilks1946}
S.~S. Wilks.
\newblock Sample criteria for testing equality of means, equality of variances,
  and equality of covariances in a normal multivariate distribution.
\newblock {\em The Annals of Mathematical Statistics}, 17(3):257--281, 1946.

\bibitem{wu1986jackknife}
C.-F.~J. Wu.
\newblock Jackknife, bootstrap and other resampling methods in regression
  analysis.
\newblock {\em The Annals of Statistics}, 14(4):1261--1295, 1986.

\bibitem{xu2013}
L.-W. Xu, F.-Q. Yang, S.~Qin, et~al.
\newblock A parametric bootstrap approach for two-way {ANOVA} in presence of
  possible interactions with unequal variances.
\newblock {\em Journal of Multivariate Analysis}, 115:172--180, 2013.

\end{thebibliography}
\bibliographystyle{abbrv}

\end{document}